\pdfoutput=1

\documentclass[12pt,a4paper]{article}

\usepackage{ifthen} 
\newboolean{pdflatex}
\setboolean{pdflatex}{true} 

\newboolean{articletitles}
\setboolean{articletitles}{true} 

\newboolean{uprightparticles}
\setboolean{uprightparticles}{false} 

\def\paperauthors{LHCb collaboration} 
\def\paperasciititle{Measurement of the Lb -> J/psi Lambda angular distribution and the Lb polarisation in pp collisions} 
\def\papertitle{Measurement of the \decay{\Lb}{\jpsi\Lz} angular distribution and the \Lb~polarisation in $pp$ collisions} 
\def\paperkeywords{{High Energy Physics}, {LHCb}} 
\def\papercopyright{\the\year\ CERN for the benefit of the LHCb collaboration} 
\def\paperlicence{CC BY 4.0 licence}
\def\paperlicenceurl{https://creativecommons.org/licenses/by/4.0/}

\usepackage[top=1in, bottom=1.25in, left=1in, right=1in]{geometry}

\usepackage{microtype}
\usepackage{lineno}  
\usepackage{xspace} 
\usepackage{caption} 

\usepackage{graphicx}  
\usepackage{color}
\usepackage{colortbl}
\graphicspath{{./figs/}} 
\DeclareGraphicsExtensions{.pdf,.PDF,png,.PNG}

\usepackage{amsmath} 
\usepackage{amssymb}
\usepackage{amsfonts}
\usepackage{upgreek} 

\newcommand*\patchAmsMathEnvironmentForLineno[1]{%
\expandafter\let\csname old#1\expandafter\endcsname\csname #1\endcsname
\expandafter\let\csname oldend#1\expandafter\endcsname\csname
end#1\endcsname
 \renewenvironment{#1}%
   {\linenomath\csname old#1\endcsname}%
   {\csname oldend#1\endcsname\endlinenomath}%
}
\newcommand*\patchBothAmsMathEnvironmentsForLineno[1]{%
  \patchAmsMathEnvironmentForLineno{#1}%
  \patchAmsMathEnvironmentForLineno{#1*}%
}
\AtBeginDocument{%
\patchBothAmsMathEnvironmentsForLineno{equation}%
\patchBothAmsMathEnvironmentsForLineno{align}%
\patchBothAmsMathEnvironmentsForLineno{flalign}%
\patchBothAmsMathEnvironmentsForLineno{alignat}%
\patchBothAmsMathEnvironmentsForLineno{gather}%
\patchBothAmsMathEnvironmentsForLineno{multline}%
\patchBothAmsMathEnvironmentsForLineno{eqnarray}%
}

\usepackage{hyperxmp}

\usepackage[pdftex,
            pdfauthor={\paperauthors},
            pdftitle={\paperasciititle},
            pdfkeywords={\paperkeywords},
            pdfcopyright={Copyright (C) \papercopyright},
            pdflicenseurl={\paperlicenceurl}]{hyperref}

\usepackage[colorinlistoftodos,textsize=scriptsize]{todonotes}

\usepackage[all]{hypcap} 

\usepackage{xspace} 
\usepackage{upgreek}

\def\lhcb   {\mbox{LHCb}\xspace}

\def\MagUp {\mbox{\em Mag\kern -0.05em Up}\xspace}

\ifthenelse{\boolean{uprightparticles}}%
{

 \def\Pmu         {\ensuremath{\upmu}\xspace}

 \def\Ppi         {\ensuremath{\uppi}\xspace}

 \def\Ppsi        {\ensuremath{\uppsi}\xspace}

 \def\PDelta      {\ensuremath{\Delta}\xspace}                 
 \def\PXi         {\ensuremath{\Xi}\xspace}                 
 \def\PLambda     {\ensuremath{\Lambda}\xspace}                 
 \def\PSigma      {\ensuremath{\Sigma}\xspace}                 
 \def\POmega      {\ensuremath{\Omega}\xspace}                 
 \def\PUpsilon    {\ensuremath{\Upsilon}\xspace}

 \def\PB      {\ensuremath{\mathrm{B}}\xspace}                 
                  
 \def\PD      {\ensuremath{\mathrm{D}}\xspace}

 \def\PJ      {\ensuremath{\mathrm{J}}\xspace}                 
 \def\PK      {\ensuremath{\mathrm{K}}\xspace}

 \def\Pb      {\ensuremath{\mathrm{b}}\xspace}                 
 \def\Pc      {\ensuremath{\mathrm{c}}\xspace}

 \def\Pi      {\ensuremath{\mathrm{i}}\xspace}

 \def\Pp      {\ensuremath{\mathrm{p}}\xspace}

 \def\Ps      {\ensuremath{\mathrm{s}}\xspace}

 \def\thebaroffset{0.0em}
}
{

 \def\Pmu         {\ensuremath{\mu}\xspace}

 \def\Ppi         {\ensuremath{\pi}\xspace}

 \def\Ppsi        {\ensuremath{\psi}\xspace}                 
                  
 \mathchardef\PDelta="7101
 \mathchardef\PXi="7104
 \mathchardef\PLambda="7103
 \mathchardef\PSigma="7106
 \mathchardef\POmega="710A
 \mathchardef\PUpsilon="7107
                  
 \def\PB      {\ensuremath{B}\xspace}                 
                  
 \def\PD      {\ensuremath{D}\xspace}

 \def\PJ      {\ensuremath{J}\xspace}                 
 \def\PK      {\ensuremath{K}\xspace}

 \def\Pb      {\ensuremath{b}\xspace}                 
 \def\Pc      {\ensuremath{c}\xspace}

 \def\Pi      {\ensuremath{i}\xspace}

 \def\Pp      {\ensuremath{p}\xspace}

 \def\Ps      {\ensuremath{s}\xspace}

 \def\thebaroffset{0.18em}
}
\newcommand{\offsetoverline}[2][\thebaroffset]{\kern #1\overline{\kern -#1 #2}}%

\makeatletter
\ifcase \@ptsize \relax
  \newcommand{\miniscule}{\@setfontsize\miniscule{4}{5}}
\or
  \newcommand{\miniscule}{\@setfontsize\miniscule{5}{6}}
\or
  \newcommand{\miniscule}{\@setfontsize\miniscule{5}{6}}
\fi
\makeatother

\DeclareRobustCommand{\optbar}[1]{\shortstack{{\miniscule (\rule[.5ex]{1.25em}{.18mm})}
  \\ [-.7ex] $#1$}}

\def\mup        {{\ensuremath{\Pmu^+}}\xspace}
\def\mun        {{\ensuremath{\Pmu^-}}\xspace} 

\def\mumu       {{\ensuremath{\Pmu^+\Pmu^-}}\xspace}

\def\squark    {{\ensuremath{\Ps}}\xspace}

\def\cquark    {{\ensuremath{\Pc}}\xspace}

\def\bquark    {{\ensuremath{\Pb}}\xspace}

\def\pion   {{\ensuremath{\Ppi}}\xspace}

\def\pip    {{\ensuremath{\pion^+}}\xspace}
\def\pim    {{\ensuremath{\pion^-}}\xspace}

\def\kaon    {{\ensuremath{\PK}}\xspace}

\def\KorKbar {\kern \thebaroffset\optbar{\kern -\thebaroffset \PK}{}\xspace}

\def\Kp      {{\ensuremath{\kaon^+}}\xspace}

\def\KS      {{\ensuremath{\kaon^0_{\mathrm{S}}}}\xspace}

\def\DorDbar {\kern \thebaroffset\optbar{\kern -\thebaroffset \PD}\xspace}

\def\B       {{\ensuremath{\PB}}\xspace}

\def\BorBbar {\kern \thebaroffset\optbar{\kern -\thebaroffset \PB}\xspace}
\def\Bz      {{\ensuremath{\B^0}}\xspace}

\def\Bd      {{\ensuremath{\B^0}}\xspace}

\def\BdorBdbar {\kern \thebaroffset\optbar{\kern -\thebaroffset \Bd}\xspace}

\def\Bs      {{\ensuremath{\B^0_\squark}}\xspace}

\def\BsorBsbar {\kern \thebaroffset\optbar{\kern -\thebaroffset \Bs}\xspace}

\def\jpsi     {{\ensuremath{{\PJ\mskip -3mu/\mskip -2mu\Ppsi\mskip 2mu}}}\xspace}

\def\Y#1S{\ensuremath{\PUpsilon{(#1S)}}\xspace}

\def\proton      {{\ensuremath{\Pp}}\xspace}
\def\antiproton  {{\ensuremath{\overline \proton}}\xspace}

\def\Lz          {{\ensuremath{\PLambda}}\xspace}
\def\Lbar        {{\ensuremath{\offsetoverline{\PLambda}}}\xspace}
\def\LorLbar     {\kern \thebaroffset\optbar{\kern -\thebaroffset \PLambda}\xspace}

\def\Sigmares    {{\ensuremath{\PSigma}}\xspace}
\def\Sigmaz      {{\ensuremath{\Sigmares{}^0}}\xspace}

\def\Xires       {{\ensuremath{\PXi}}\xspace}

\def\Lb           {{\ensuremath{\Lz^0_\bquark}}\xspace}
\def\Lbbar        {{\ensuremath{\Lbar{}^0_\bquark}}\xspace}

\newcommand{\decay}[2]{\ensuremath{#1\!\to #2}\xspace} 

\def\to                 {\ensuremath{\rightarrow}\xspace}

\def\CP                {{\ensuremath{C\!P}}\xspace}

\def\AT#1     {\ensuremath{A_{\mathrm{T}}^{#1}}\xspace}           

\def\C#1      {\ensuremath{\mathcal{C}_{#1}}\xspace}                       
\def\Cp#1     {\ensuremath{\mathcal{C}_{#1}^{'}}\xspace}                    
\def\Ceff#1   {\ensuremath{\mathcal{C}_{#1}^{\mathrm{(eff)}}}\xspace}        
\def\Cpeff#1  {\ensuremath{\mathcal{C}_{#1}^{'\mathrm{(eff)}}}\xspace}       
\def\Ope#1    {\ensuremath{\mathcal{O}_{#1}}\xspace}                       
\def\Opep#1   {\ensuremath{\mathcal{O}_{#1}^{'}}\xspace}                    

\newcommand{\nospaceunit}[1]{\ensuremath{\text{#1}}}       
\newcommand{\aunit}[1]{\ensuremath{\text{\,#1}}}       

\newcommand{\tev}{\aunit{Te\kern -0.1em V}\xspace}
\newcommand{\gev}{\aunit{Ge\kern -0.1em V}\xspace}
\newcommand{\mev}{\aunit{Me\kern -0.1em V}\xspace}
\newcommand{\kev}{\aunit{ke\kern -0.1em V}\xspace}
\newcommand{\ev}{\aunit{e\kern -0.1em V}\xspace}
\newcommand{\mevc}{\ensuremath{\aunit{Me\kern -0.1em V\!/}c}\xspace}
\newcommand{\gevc}{\ensuremath{\aunit{Ge\kern -0.1em V\!/}c}\xspace}
\newcommand{\mevcc}{\ensuremath{\aunit{Me\kern -0.1em V\!/}c^2}\xspace}
\newcommand{\gevcc}{\ensuremath{\aunit{Ge\kern -0.1em V\!/}c^2}\xspace}

\def\mm   {\aunit{mm}\xspace}

\def\mum  {\ensuremath{\,\upmu\nospaceunit{m}}\xspace}

\def\fb   {\ensuremath{\aunit{fb}}\xspace}
\def\invfb   {\ensuremath{\fb^{-1}}\xspace}

\def\ps   {\ensuremath{\aunit{ps}}\xspace}

\newcommand{\chisq}{\ensuremath{\chi^2}\xspace}

\newcommand{\chisqip}{\ensuremath{\chi^2_{\text{IP}}}\xspace}

\def\deriv {\ensuremath{\mathrm{d}}}

\def\gsim{{~\raise.15em\hbox{$>$}\kern-.85em
          \lower.35em\hbox{$\sim$}~}\xspace}
\def\lsim{{~\raise.15em\hbox{$<$}\kern-.85em
          \lower.35em\hbox{$\sim$}~}\xspace}

\def\sPlot{\mbox{\em sPlot}\xspace}

\def\pt         {\ensuremath{p_{\mathrm{T}}}\xspace}

\def\ptot       {\ensuremath{p}\xspace}

\def\mrad{\aunit{mrad}}

\def\evtgen     {\mbox{\textsc{EvtGen}}\xspace}

\def\geant      {\mbox{\textsc{Geant4}}\xspace}

\def\photos     {\mbox{\textsc{Photos}}\xspace}

\def\pythia     {\mbox{\textsc{Pythia}}\xspace}

\def\tell1  {TELL1\xspace}
\def\ukl1   {UKL1\xspace}

\newcommand{\cf}{\mbox{\itshape cf.}\xspace}

\usepackage{cite} 
\usepackage{mciteplus}

\usepackage{booktabs} 

\begin{document}

\renewcommand{\thefootnote}{\fnsymbol{footnote}}
\setcounter{footnote}{1}


\begin{titlepage}
\pagenumbering{roman}

\vspace*{-1.5cm}
\centerline{\large EUROPEAN ORGANIZATION FOR NUCLEAR RESEARCH (CERN)}
\vspace*{1.5cm}
\noindent
\begin{tabular*}{\linewidth}{lc@{\extracolsep{\fill}}r@{\extracolsep{0pt}}}
\ifthenelse{\boolean{pdflatex}}
{\vspace*{-1.5cm}\mbox{\!\!\!\includegraphics[width=.14\textwidth]{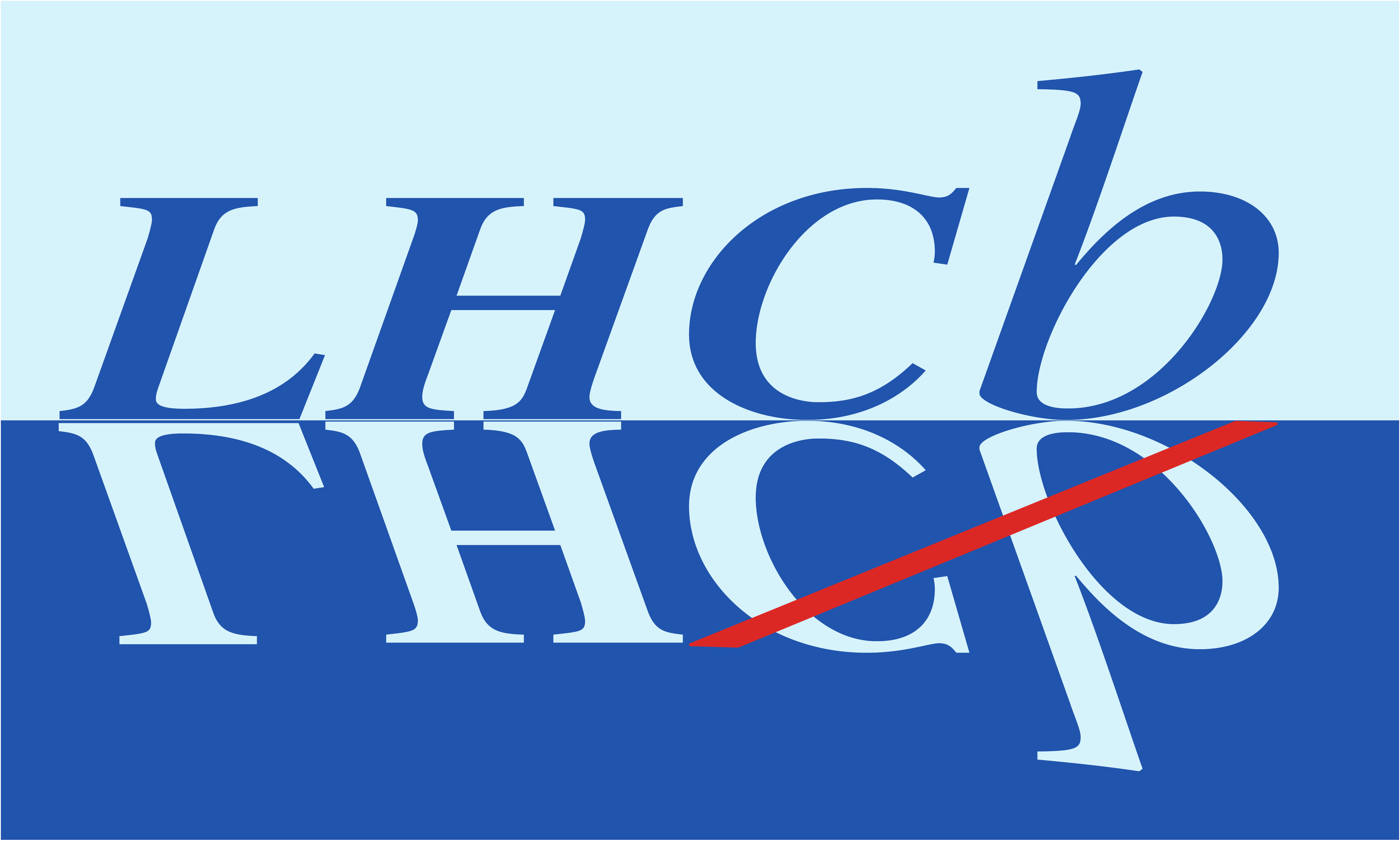}} & &}%
{\vspace*{-1.2cm}\mbox{\!\!\!\includegraphics[width=.12\textwidth]{figs/lhcb-logo.eps}} & &}%
\\
 & & CERN-EP-2020-051 \\  
 & & LHCb-PAPER-2020-005 \\  
 & & 22 April 2020 \\ 
 & & \\
\end{tabular*}

\vspace*{4.0cm}

{\normalfont\bfseries\boldmath\huge
\begin{center}
  \papertitle 
\end{center}
}

\vspace*{2.0cm}

\begin{center}
\paperauthors\footnote{Authors are listed at the end of this paper.}
\end{center}

\vspace{\fill}

\begin{abstract}
  \noindent
  This paper presents an analysis of the \decay{\Lb}{\jpsi\Lz} angular distribution and the transverse production polarisation of \Lb baryons in proton-proton collisions at centre-of-mass energies of 7, 8 and 13\tev. The measurements are performed using data corresponding to an integrated luminosity of 4.9\invfb, collected with the LHCb experiment.
  The polarisation is determined in a fiducial region of \Lb transverse momentum and pseudorapidity of $1 < \pt < 20\gevc$ and $2 < \eta < 5$, respectively. 
  The data are consistent with \Lb baryons being produced unpolarised in this region.
  The parity-violating asymmetry parameter of the \decay{\Lz}{p\pim} decay is also determined from the data and its value is found to be consistent with a recent measurement by the BES\,III collaboration. 
\end{abstract}

\vspace*{2.0cm}

\begin{center}
Published in J. High Energ. Phys. 2020, 110 (2020)
\end{center}

\vspace{\fill}

{\footnotesize 
\centerline{\copyright~\papercopyright. \href{\paperlicenceurl}{\paperlicence}.}}
\vspace*{2mm}

\end{titlepage}


\newpage
\setcounter{page}{2}
\mbox{~}
%

\cleardoublepage


\renewcommand{\thefootnote}{\arabic{footnote}}
\setcounter{footnote}{0}


\pagestyle{plain} 
\setcounter{page}{1}
\pagenumbering{arabic}


\section{Introduction}
\label{sec:Introduction}
 
Studies of the production and decay of heavy-flavour hadrons are an important part of contemporary particle physics.
The spin-$\tfrac{1}{2}$ \Lb baryon can provide information about the production of hadrons containing \bquark quarks.
For example, the \Lb polarisation is closely related to that of the \bquark quark~\cite{Hiller:2007ur}. 
Heavy-quark effective theory (HQET) predicts that \Lb baryons originating from energetic \bquark quarks retain a large fraction of the transverse \bquark-quark polarisation \cite{Mannel:1991bs,Falk:1993rf}.
The longitudinal polarisation is expected to vanish in $pp$ collisions due to parity conservation in strong interactions and the term polarisation is used to refer to the transverse polarisation of particles in this paper. 
The authors of Ref.~\cite{Dharmaratna:1996xd} estimate that the \bquark-quark polarisation is of the order of 10\%.
This leads to an estimate that the polarisation of the \Lb baryon can be around 10\% with possible values up to 20\% \cite{Ajaltouni:2004zu, Hiller:2007ur}.
Measurements of \Lz polarisation at fixed-target experiments \cite{Ramberg:1994tk,Fanti:1998px,Abt:2006da} find that the polarisation strongly depends on Feynman-$x$, $x_{\rm F}$, with polarisation vanishing at $x_{\rm F}=0$. The variable $x_{\rm F}$ is defined by $x_{\rm F}=2p_{\rm L}/\sqrt{s}$, where $p_{\rm L}$ is the longitudinal momentum of the baryon with respect to the beam line and $\sqrt{s}$ is the centre-of-mass energy of the collision.
If a similar $x_{\rm F}$-dependence is present in \Lb-baryon production, a negligible polarisation would be expected at the LHC since the experiments mostly cover the phase-space region close to $x_{\rm F}=0$.
In addition, several heavy \bquark-baryon states are observed experimentally~\cite{Aaltonen:2007ar,LHCb-PAPER-2012-012,Aaltonen:2013tta,LHCb-PAPER-2019-025,Sirunyan:2020gtz}.
In the production of \Lb baryons from decays of these states, the  connection between the \Lb and the \bquark-quark polarisation can be further diluted due to the interaction of the \bquark quark with the light quarks in the heavy \bquark-baryon~\cite{Hiller:2007ur,Falk:1993rf}.
The fraction of the \bquark-quark polarisation transferred to the \Lb baryon is estimated to be around 75\% in Ref.~\cite{Hiller:2007ur}.

The decay \decay{\Lb}{\jpsi\Lz}, where the \Lz baryon decays to $\proton\pim$ and the \jpsi meson decays to \mumu, can be used to measure the polarisation of the \Lb baryon as well as to test the theoretical understanding of hadronic decays of \Lb baryons.\footnote{The inclusion of charge-conjugate processes is implied throughout this paper except when stated otherwise.}
The angular distribution of the \decay{\Lb}{\jpsi\Lz} decay is described by the polarisation of the \Lb baryon, $P_{b}$, four decay amplitudes and by the parity-violating asymmetry parameter of the \Lz baryon decay, $\alpha_\Lz$.
The decay parameter $\alpha_\Lz$ arises due to the V$-$A nature of the weak interaction~\cite{Lee:1957qs}. 
The four decay amplitudes, $A(\lambda_{\Lz},\lambda_{\jpsi})$ correspond to different \Lz  and \jpsi helicities, $\lambda_{\Lz}$ and $\lambda_{\jpsi}$. The  notation $a_{\pm} = A(\pm\tfrac{1}{2},0)$ and $b_{\pm} = A(\mp\tfrac{1}{2},\pm 1)$ is used in this paper.

In the naive heavy-quark and light-diquark limit, the $u$ and $d$ quark in the baryon form a spin- and isospin-zero spectator system. 
The left-handed nature of the charged-current interaction then implies that the \Lz-baryon helicity is $-\tfrac{1}{2}$, such that $|a_+| \approx |b_-| \approx 0$. 
Several theoretical approaches have been used to predict the \Lb parity-violating decay parameter
\begin{align}
\alpha_b = \frac{|a_+|^{2} -  |a_-|^{2} + |b_+|^{2} -  |b_-|^{2}}{|a_+|^{2} +  |a_-|^{2} + |b_+|^{2} +  |b_-|^{2}} ~,
\end{align}
which is the analogue of $\alpha_{\Lz}$ but applied to the \Lb decay. 
The value of $\alpha_b$ is predicted to be in the range from $-0.2$ to $-0.1$ within a factorisation approximation~\cite{Cheng:1996cs,Fayyazuddin:1998ap,Fayyazuddin:2017sxq}, around $-0.2$ in the covariant oscillator quark model \cite{Mohanta:1998iu} or light-front quark model \cite{Wei:2009np} and in the range from $-0.17$ to $-0.14$ in approaches based on perturbative QCD \cite{Chou:2001bn}.
In contrast, a prediction based on HQET yields a value of $\alpha_b \sim 0.8$~\cite{Ajaltouni:2004zu}.
The covariant quark model has recently been used to predict $\alpha_b \sim -0.07$ and the  magnitudes of the four helicity amplitudes~\cite{Gutsche:2018utw,Gutsche:2017wag}. 
The amplitudes predicted by this model agree with the naive expectation that $|a_+|$ and $|b_-|$ are small, while $|a_-|$ and $|b_+|$ are of similar size.

The polarisation of \Lb baryons was previously measured at LEP in $Z$ decays \cite{Buskulic:1995mf,Abbiendi:1998uz,Abreu:1999gf} and at the LHC in $pp$ collisions~\cite{LHCb-PAPER-2012-057,Sirunyan:2018bfd}.
The values measured at the LHC are
\begin{displaymath}
\begin{split}
P_b & = 0.06\pm 0.07 \pm 0.02 \quad({\rm LHCb})~, \\
P_b & = 0.00\pm 0.06 \pm 0.02 \quad({\rm CMS}) ~.
\end{split}
\end{displaymath}
Both measurements were performed using an angular analysis of the \decay{\Lb}{\jpsi\Lz} decay. 
The LHCb measurement used data collected at $\sqrt{s}=7\tev$, while  
the CMS measurement used data from both 7 and 8\tev $pp$ collisions.
A similar analysis was performed by the ATLAS collaboration \cite{Aad:2014iba} but assuming $P_b=0$ and measuring only magnitudes of the decay amplitudes.
While all three measurements are compatible, the LHCb and CMS results are unphysical; the LHCb value of $|b_{-}|^{2}$ and the CMS value of $|a_{+}|^{2}$ are  negative. 
This is likely to be due to the use of a now outdated value of $\alpha_\Lz=0.642\pm0.013$ from an earlier Particle Data Group  average of the results of  Refs.~\cite{Astbury:1975hn,Cleland:1972fa,Dauber:1969hg,Overseth:1967zz,Cronin:1963zb} that is no longer used. 
This value is significantly smaller than that measured by the BES\,III collaboration using \decay{\jpsi}{\Lz\Lbar} decays~\cite{Ablikim:2018zay}. 
In their analysis, the BES\,III collaboration determine $\alpha_\Lz$ and $\alpha_\Lbar$, for the \decay{\Lz}{\proton\pim} and \decay{\Lbar}{\antiproton\pip} decays, to be $\alpha_{\Lz} = 0.750\pm 0.009 \pm 0.004$ and $\alpha_{\Lbar} = -0.758\pm 0.010\pm 0.007$.
The BES\,III measurement is supported by a reanalysis of CLAS $\gamma\proton\to\Kp\Lz$ scattering data in Ref.~\cite{Ireland:2019uja}, which gives $\alpha_\Lz = 0.721\pm0.006\pm0.005$.
The polarisation of \Lb baryons has also been  determined to be $P_{b} = (0\pm5)\%$ in the LHCb acceptance using \decay{\Lb}{\Lz\mumu} decays, under the assumption that the polarisation is independent of $\sqrt{s}$~\cite{Blake:2019guk}. 

This paper describes a measurement of the \decay{\Lb}{\jpsi\Lz} angular distribution using data collected with the LHCb experiment during Run\,1 and Run\,2 of the LHC. 
The data set corresponds to 1.0, 2.0 and 1.9\invfb of integrated luminosity collected at $\sqrt{s} = 7$, 8 and 13\tev in 2011, 2012 and 2015--2016, respectively. 
A measurement of the polarisation and the decay amplitudes is 
made, using the BES\,III value of $\alpha_{\Lz}$ as an input.
The polarisation of \Lb baryons is measured for the first time at $\sqrt{s}=13\tev$. 

The paper starts by describing the angular formalism used in the analysis in Section~\ref{sec:formalism}.
Section~\ref{sec:Detector} introduces the LHCb detector. 
Section~\ref{sec:Selection} describes the selection of candidates from the LHCb data set. 
The yields of \decay{\Lb}{\jpsi\Lz} decays in the different data sets are obtained in Section~\ref{sec:Yields}.
Section~\ref{sec:Efficiency} describes the procedure used to correct the data for the nonuniformity of the reconstruction and selection. 
The production polarisation and decay amplitudes are obtained through a two-step procedure described in Sections~\ref{sec:Moments} and \ref{sec:Results}. 
Section~\ref{sec:Systematics} discusses sources of systematic uncertainty in the measurement. 
Finally, conclusions are presented in Section~\ref{sec:Summary}.

\section{Angular formalism}
\label{sec:formalism} 

\begin{figure}[!tb]
\centering
\includegraphics[width=0.7\linewidth]{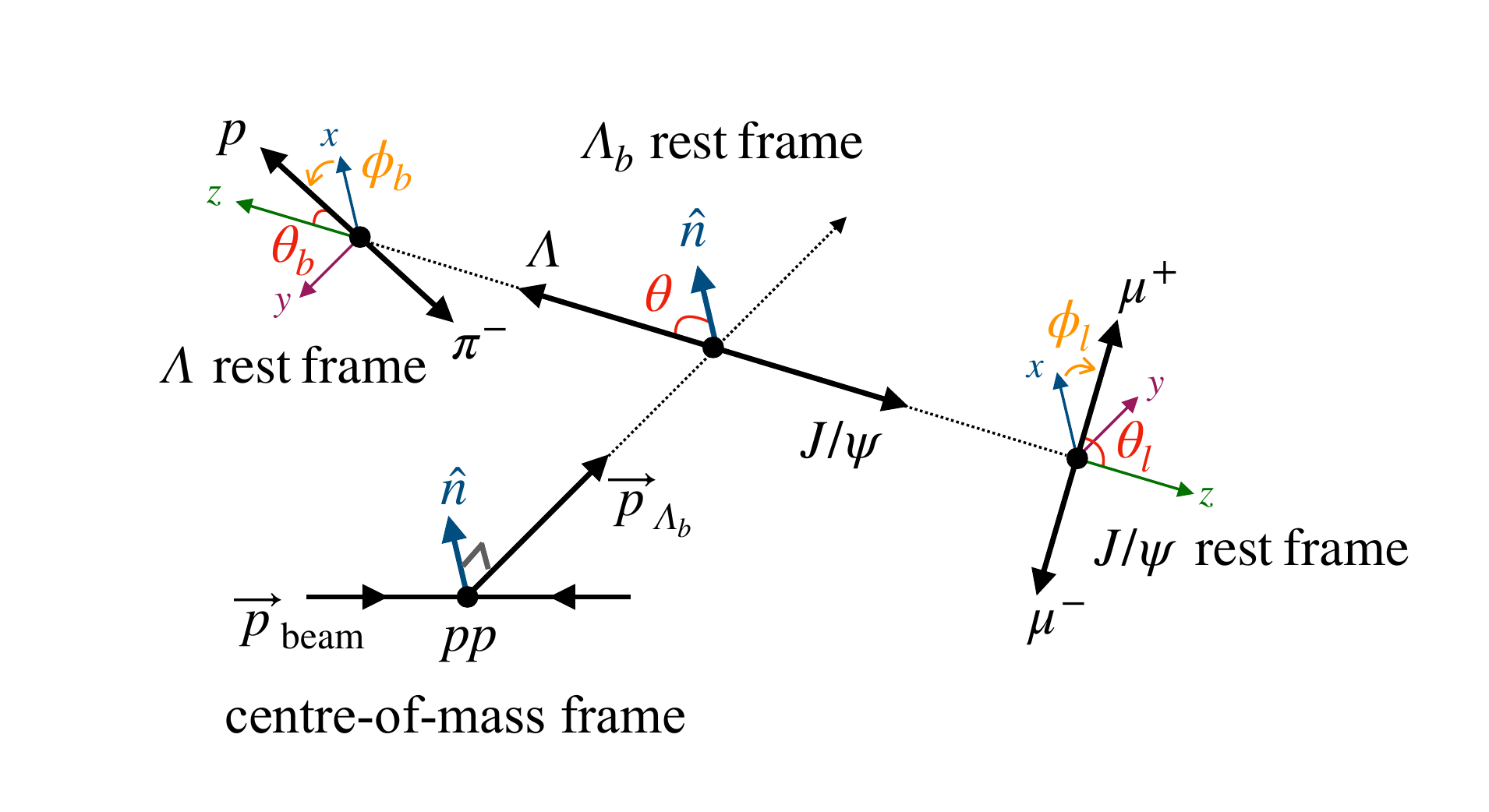}
\caption{
Definition of the five decay angles, $\theta$, $\theta_b$, $\phi_b$, $\theta_{l}$ and $\phi_l$ used to describe the kinematics of the \decay{\Lb}{\jpsi\Lz} decay. The angles are described in the text.
}
\label{fig:basis}
\end{figure}

The kinematics of the \decay{\Lb}{\jpsi\Lz} decay, including the subsequent decays of the \jpsi meson and the \Lz baryon,  can be parameterised by five decay angles and a unit vector in the direction transverse to the production plane, $\hat{n}$, against which the polarisation is measured~\cite{Blake:2017une}. 
The unit vector is defined as \mbox{$\hat{n} = (\vec{p}_{\rm beam} \times \vec{p}_{\Lb})/|\vec{p}_{\rm beam} \times \vec{p}_{\Lb}|$}, where $\vec{p}_{\Lb}$ and $\vec{p}_{\rm beam}$ are vectors in the direction of the \Lb baryon and the beam in the centre-of-mass frame of the $pp$ collision. 
In the case of the LHCb detector, $\vec{p}_{\rm beam}$ is the direction of the beam that points into the detector from the collision point. 
The four-momentum of each particle is boosted into the centre-of-mass frame to account for the small beam-crossing angle of the LHC collisions before $\hat{n}$ is calculated. 
The five decay angles are:
the angle, $\theta$, between $\hat{n}$ and the \Lz flight direction in the \Lb rest frame;
the polar, $\theta_{b}$, and azimuthal, $\phi_b$, angles of the proton in the \Lz rest frame;
and the polar, $\theta_{l}$, and azimuthal, $\phi_{l}$, angles of the \mup in the \jpsi rest frame. 
The angles $\theta$, $\theta_l$ and $\theta_b$ are defined in the range $[0,\pi]$, while $\phi_l$ and $\phi_b$ are defined over $[-\pi,+\pi]$.
A visual depiction of the angular basis is given in Fig.~\ref{fig:basis}. 
The decay angles for the \Lbbar decay are defined assuming no \CP violation in the \Lb or \Lz decay,  such that the distributions of \Lb and \Lbbar decays are identical. 

The angular distribution of the \decay{\Lb}{\jpsi\Lz} decay can be expressed as~\cite{Hrivnac:1994jx} 
\begin{align}
    \frac{\deriv^{5}\Gamma}{\deriv\vec{\Omega}} &= \frac{3}{32\pi^{2}} \sum\limits_{i} J_{i}(a_{+},a_{-},b_{+},b_{-},\alpha_{\Lz},P_{b}) f_{i}(\vec{\Omega})~,
    \label{eq:angular}
\end{align}
where $\vec{\Omega} = (\cos\theta, \cos\theta_{b}, \phi_{b}, \cos\theta_{l}, \phi_{l})$. 
The angular terms, $J_{i}$, and the angular functions, $f_{i}$, are given in Table~\ref{tab:angular}. 
The \Lb polarisation is accessible through terms  $J_{11}$--$J_{34}$. 

\begin{table}[!tb]
    \caption{
    Angular functions parameterising the \decay{\Lb}{\jpsi\Lz} angular distribution. 
    The numbering scheme is the same as in Ref.~\cite{Blake:2017une}. 
    }
    \centering
    {\renewcommand{\arraystretch}{1.5} 
    \begin{tabular}{c|c|c}
    \toprule 
    $i$ & $J_{i}$ & $f_{i}(\vec{\Omega})$ \\
    \midrule
    1 & $\tfrac{1}{4}(2 |a_{+}|^2 + 2|a_{-}|^{2} + |b_{+}|^{2} + |b_{-}|^{2})$ & $\sin^{2}\theta_{l}$\\
    2 & $\tfrac{1}{2} |b_{+}|^{2} + \tfrac{1}{2} |b_{-}|^{2}$ & $\cos^2\theta_l$ \\
    4 & $ \tfrac{1}{4} \alpha_{\Lz}(2|a_+|^2-2|a_-|^2 -|b_+|^2 + |b_-|^2)$ & $\sin^{2}\theta_{l}\cos\theta_{b}$ \\
    5 &  $\tfrac{1}{2} \alpha_{\Lz} (|b_-|^2-|b_+|^2)$ & $\cos^{2}\theta_{l} \cos\theta_{b}$ \\
    7 &  $\tfrac{1}{\sqrt{2}} \alpha_{\Lz} {\rm Re}(-b_+^*a_++b_-a_-^*)$ & $\sin\theta_l\cos\theta_l\sin\theta_b\cos\left(\phi_b+\phi_l\right)$ \\
    9 &  $\tfrac{1}{\sqrt{2}} \alpha_{\Lz} {\rm Im}(b_+^*a_+-b_-a_-^*)$  & $\sin\theta_l\cos\theta_l\sin\theta_b\sin\left(\phi_b+\phi_l\right)$\\
    11 &  $\tfrac{1}{4} P_b (2|a_+|^2 - 2|a_-|^2 + |b_+|^2 - |b_-|^2)$ & $\sin^2\theta_l \cos\theta$ \\
    12 &  $ \tfrac{1}{2} P_b(|b_+|^2-|b_-|^2)$ & $\cos^2\theta_l \cos\theta$ \\
    14 &  $\tfrac{1}{4} P_b \alpha_{\Lz} (2|a_+|^2+2|a_-|^2 -|b_+|^2-|b_-|^2)$ & $\sin^{2}\theta_{l}\cos\theta_{b}\cos\theta$\\
    15 &  $-\tfrac{1}{2} P_b \alpha_{\Lz} (|b_+|^2+|b_-|^2)$ & $\cos^{2}\theta_{l} \cos\theta_{b}\cos\theta$\\
    17 &  $-\tfrac{1}{\sqrt{2}} P_b \alpha_{\Lz} {\rm Re}(b_+^*a_++b_-a_-^*)$ & $\sin\theta_l\cos\theta_l\sin\theta_b\cos\left(\phi_b+\phi_l\right)\cos\theta$\\
    19 &  $\tfrac{1}{\sqrt{2}} P_b \alpha_{\Lz} {\rm Im}(b_+^*a_++b_-a_-^*)$  & $\sin\theta_l\cos\theta_l\sin\theta_b\sin\left(\phi_b+\phi_l\right)\cos\theta$\\
    21 &  $-\tfrac{1}{\sqrt{2}}P_b {\rm Im}(b_+^*a_--b_-a_+^*)$ & $\sin\theta_l\cos\theta_l\sin\phi_{l}\sin\theta$ \\
    23 &  $\tfrac{1}{\sqrt{2}} P_b {\rm Re}(b_+^*a_--b_-a_+^*)$ & $\sin\theta_l\cos\theta_l\cos\phi_{l}\sin\theta$\\
    25 &  $\tfrac{1}{\sqrt{2}} P_b \alpha_{\Lz}{\rm Im}(b_+^*a_-+b_-a_+^*)$ & $\sin\theta_l\cos\theta_l\cos\theta_{b}\sin\phi_{l}\sin\theta$ \\
    27 &  $-\tfrac{1}{\sqrt{2}} P_b \alpha_{\Lz} {\rm Re}(b_+^*a_-+b_-a_+^*)$ & $\sin\theta_l\cos\theta_l\cos\theta_{b}\cos\phi_{l}\sin\theta$ \\
    30 &  $P_b \alpha_{\Lz}{\rm Im}(a_+a_-^*)$ & $\sin^{2}\theta_{l} \sin\theta_{b}\sin\phi_{b} \sin\theta$ \\
    32 &  $-P_b \alpha_{\Lz} {\rm Re}(a_+a_-^*)$ & $\sin^{2}\theta_{l} \sin\theta_{b}\cos\phi_{b} \sin\theta$ \\
    33 &  $-\tfrac{1}{2} P_b \alpha_{\Lz} {\rm Re}(b_+^*b_-)$ & $\sin^{2}\theta_{l} \sin\theta_{b} \cos( 2\phi_{l} + \phi_{b}) \sin\theta$\\
    34 &  $\tfrac{1}{2} P_b \alpha_{\Lz} {\rm Im}(b_+^*b_-)$ & $\sin^{2}\theta_{l} \sin\theta_{b} \sin( 2\phi_{l} + \phi_{b}) \sin\theta$\\ 
    \bottomrule 
    \end{tabular}
    }
    \label{tab:angular}
\end{table}

\section{Detector and simulation}
\label{sec:Detector}

The \lhcb detector~\cite{LHCb-DP-2008-001,LHCb-DP-2014-002} is a single-arm forward
spectrometer covering the pseudorapidity range $2<\eta <5$,
designed for the study of particles containing \bquark or \cquark
quarks. The detector includes a high-precision tracking system
consisting of a silicon-strip vertex detector surrounding the $pp$
interaction region~\cite{LHCb-DP-2014-001}, a large-area silicon-strip detector located
upstream of a dipole magnet with a bending power of about
$4{\mathrm{\,Tm}}$, and three stations of silicon-strip detectors and straw
drift tubes~\cite{LHCb-DP-2013-003,LHCb-DP-2017-001} placed downstream of the magnet.
The tracking system provides a measurement of the momentum, \ptot, of charged particles with
a relative uncertainty that varies from 0.5\% at low momentum to 1.0\% at 200\gevc.
The minimum distance of a track to a primary $pp$ collision vertex (PV), the impact parameter (IP), 
is measured with a resolution of $(15+(29\gevc)/\pt)\mum$,
where \pt is the component of the momentum transverse to the beam. 
Muons are identified by a system composed of alternating layers of iron and multiwire
proportional chambers~\cite{LHCb-DP-2012-002}.
The online event selection is performed by a trigger~\cite{LHCb-DP-2012-004}, 
which consists of a hardware stage, based on information from the muon
system and calorimeters, followed by a software stage, which applies a full event
reconstruction.

Samples of simulated events are required to model the effects of the detector acceptance and the
imposed selection requirements on the \decay{\Lb}{\jpsi\Lz} angular distribution.
In the simulation, $pp$ collisions are generated using
\pythia~\cite{Sjostrand:2007gs} with a specific \lhcb
configuration~\cite{LHCb-PROC-2010-056}.  Decays of unstable particles
are described by \evtgen~\cite{Lange:2001uf}, in which final-state
radiation is generated using \photos~\cite{Golonka:2005pn}. The
interaction of the generated particles with the detector, and its response,
are implemented using the \geant
toolkit~\cite{Allison:2006ve, *Agostinelli:2002hh} as described in
Ref.~\cite{LHCb-PROC-2011-006}.
The \pt distribution of the simulated \Lb baryons is weighted to match the spectrum observed in Ref.~\cite{LHCb-PAPER-2014-004}.

\section{Candidate selection}
\label{sec:Selection}

Signal candidates are formed by combining a \jpsi-meson candidate with a \Lz-baryon candidate. 
The \jpsi candidates are reconstructed from two oppositely charged tracks that have been identified as muons. 
The muons are required to have a significant IP with respect to all PVs in the event and form a common vertex with a good vertex-fit quality. 
The dimuon mass is required to be in the range $2900 < m(\mumu) < 3150 \mevcc$.
The \Lz candidates are reconstructed in two categories: 
\Lz baryons that decay early enough for the proton and pion to be reconstructed in the vertex detector; 
and \Lz baryons that decay later, such that they cannot be reconstructed in the vertex detector. 
These categories are referred to as long and downstream, respectively. The \Lz candidates in the 
long category have a better mass, momentum and vertex resolution than those in the 
downstream category.
Approximately two thirds of the candidates are reconstructed in the downstream category. 
For both categories, the proton and pion are required to be significantly displaced from all PVs in the event and form a common vertex with a good vertex-fit quality. 
The \Lz candidates are also required to have an invariant mass within 30\mevcc of the known \Lz-baryon mass~\cite{PDG2018}, a decay time larger than 2\ps and a decay vertex at $z < 2350\mm$. The $z$-axis is aligned with the LHC beam line, with positive $z$ in the direction of the LHCb detector acceptance, where $z = 0$ corresponds approximately to the centre of the $pp$ interaction region.
The vertex position requirement is imposed to remove background from material interactions in front of the large-area silicon-strip detector. 
The \Lb candidate is associated with the PV relative to which it has smallest \chisqip, where \chisqip is defined as the difference in the vertex-fit \chisq of a given PV reconstructed with and without a considered particle.
The \Lb candidate is required to have a good vertex-fit quality, to be consistent with originating from its associated PV and to have a vertex position that is significantly displaced from that PV.  
A kinematic fit is then performed, constraining the masses of the \jpsi and \Lz candidates to their known values~\cite{PDG2018} and constraining the \Lb candidate to originate from its associated PV.

The signal candidates are required to have passed a hardware trigger that selects either a single muon with a large transverse momentum or a pair of muons with a large product of their individual transverse momenta. 
The software trigger requires a candidate to be at least partially reconstructed with a secondary vertex that has a significant displacement from any PV. 
At least one charged particle must have a large \pt and be inconsistent with originating from a PV.
A multivariate discriminator~\cite{BBDT} is used for the identification of secondary vertices consistent with the decay of a \bquark hadron.

A neural network~\cite{feindt-2004,Feindt:2006pm} is trained to reject background from events where tracks have been mistakenly combined to form a signal candidate (combinatorial background).
The network is trained using simulated \decay{\Lb}{\jpsi\Lz} decays as a signal sample and candidates from the data with a $\jpsi\Lz$ invariant mass, $m({\jpsi\Lz})$, larger than $5900\mevcc$ as a background sample. 
The neural network uses the following inputs: 
the \Lb decay time and \pt;
the \Lz mass, decay time and \pt; 
the \chisq of the fitted \Lb decay vertex; 
the angle between the \Lb momentum direction and the vector connecting the primary and \Lb decay vertices;  
and the \chisqip of the final-state hadron and muon with the largest \pt with respect to its associated PV.  
Separate classifiers are trained for data taken at different collision energies. 
A single neural network is used for both long and downstream candidates, with the \Lz category used as an input to the network.  
The working point of the neural network is chosen to maximise $\varepsilon_{S} S/\sqrt{\varepsilon_{S} S+\varepsilon_{B} B}$. Here, $S$ and $B$ are the number of signal and background decays within 14\mevcc of the known \Lb mass~\cite{PDG2018} (about twice the resolution on the invariant mass) before the application of the classifier, $\varepsilon_{S}$ and $\varepsilon_{B}$ are the efficiencies of the classifier requirement evaluated on the signal and background training samples. 

The \Lb candidates are required to be in the fiducial region, $1 < \pt < 20\gevc$ and $2 < \eta < 5$.
The mean of the $x_{\rm F}$ distribution of the selected \Lb signal decays varies  between 0.015 at $\sqrt{s}=13\tev$ and 0.028 at 7\tev. 
The corresponding standard deviations of these distributions are 0.008 and 0.014. 

Several sources of specific background have been considered. 
The largest specific background originates from \decay{\Bz}{\jpsi\KS} decays, where one of the pions from the \decay{\KS}{\pip\pim} decay is reconstructed as a proton. 
Background from partially reconstructed \bquark-baryon decays such as \decay{\Lb}{\jpsi\Lz(1520)}, 
\decay{\Lb}{\jpsi\Sigmaz} or \decay{\Xires_b}{\jpsi\Xires} decays, where the $\Lz(1520)$, $\Sigmaz$ and $\Xires$ subsequently decay to a \Lz baryon, give a negligible contribution to the selected sample. 

\section{Signal yields}
\label{sec:Yields}

\begin{figure}
    \centering
    \includegraphics[width=0.48\linewidth]{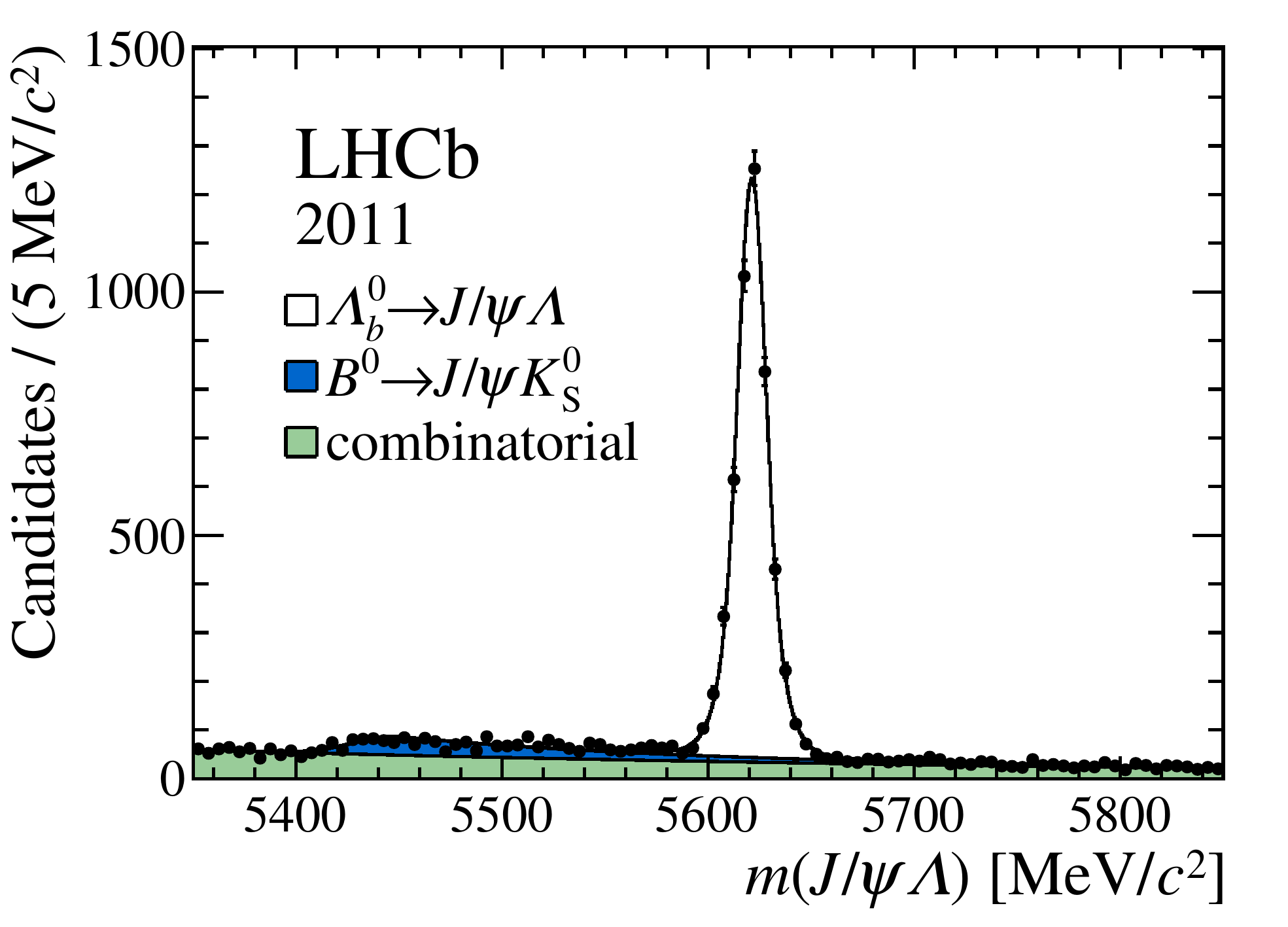}
    \includegraphics[width=0.48\linewidth]{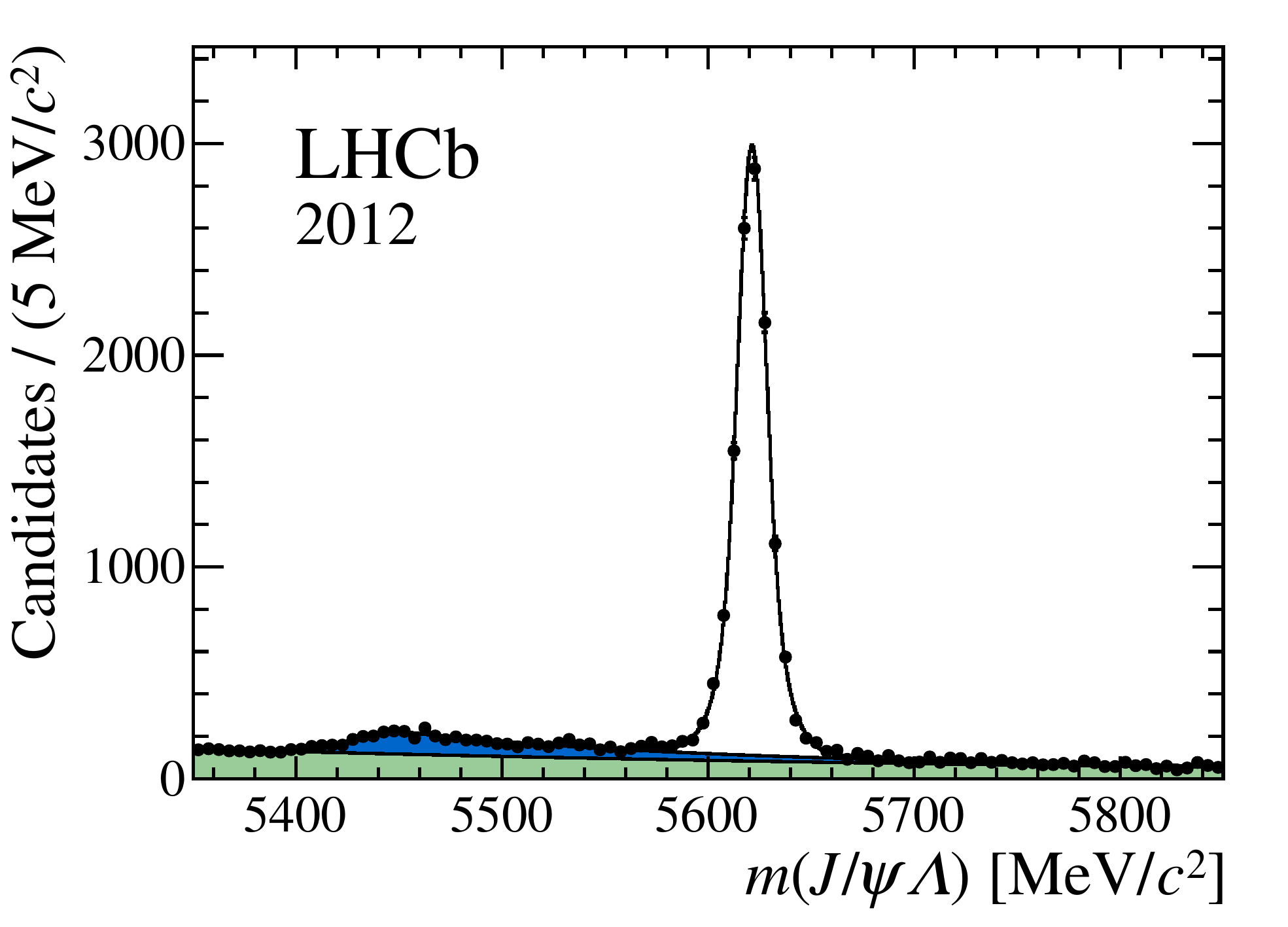} \\
    \includegraphics[width=0.48\linewidth]{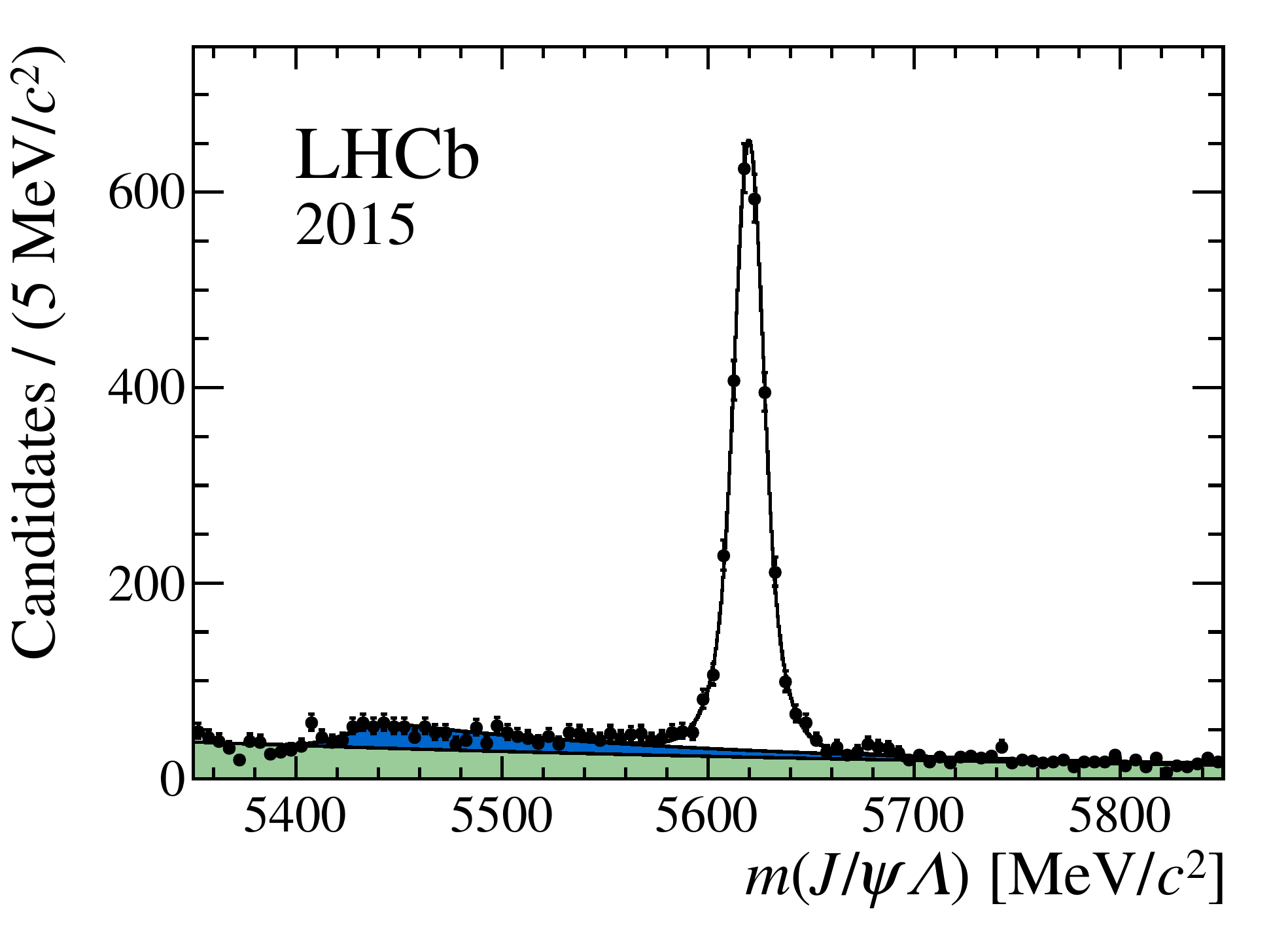}
    \includegraphics[width=0.48\linewidth]{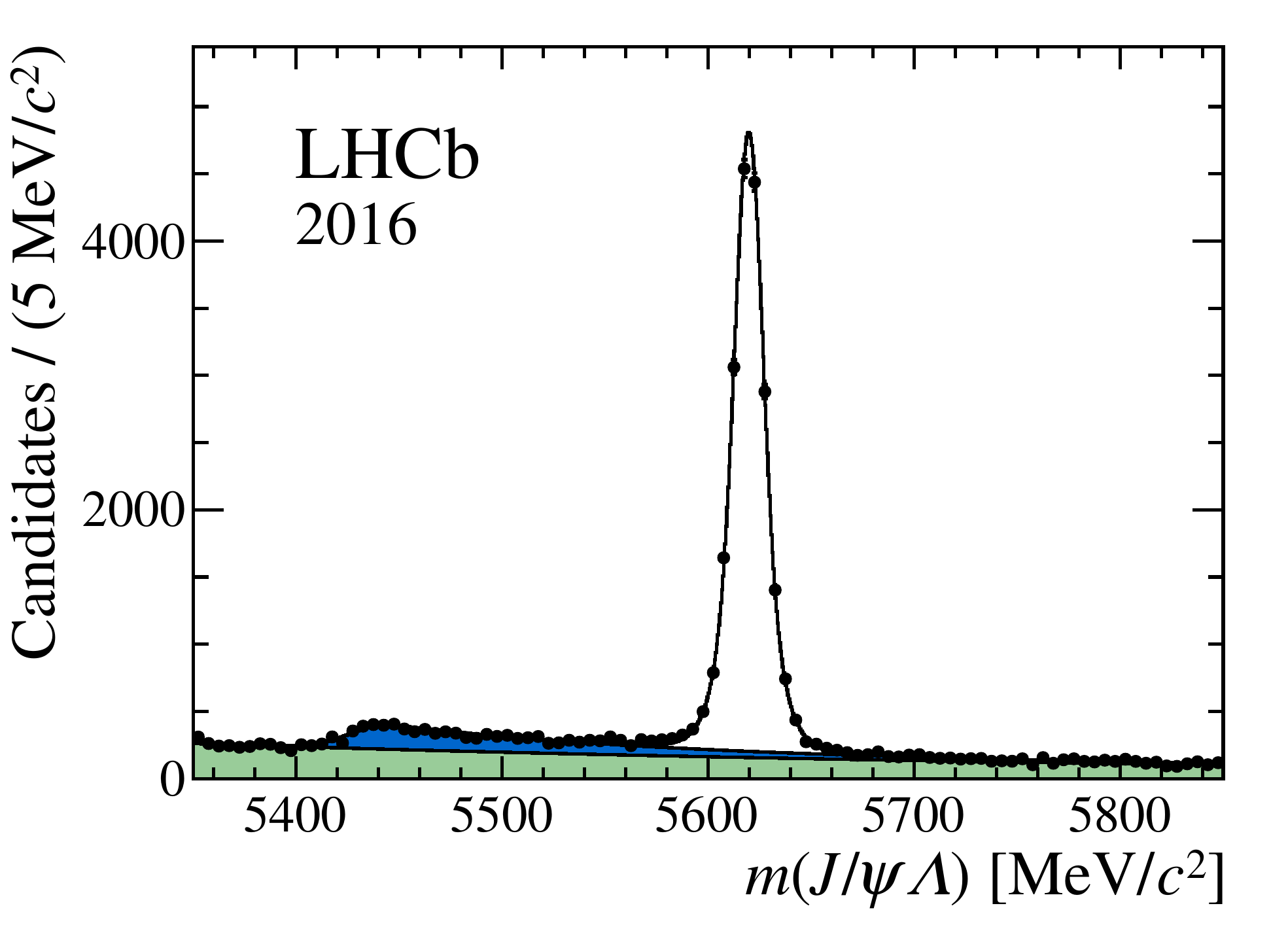}
    \caption{
    Mass distribution of selected \decay{\Lb}{\jpsi\Lz} candidates in (top-left) the 2011, (top-right) the 2012, (bottom-left) the 2015 and (bottom-right) the 2016 data sets. 
    The long and downstream categories have been combined.
    The results of fits to the distributions are overlaid.
    }
    \label{fig:selection:mass}
\end{figure}

The yield of \decay{\Lb}{\jpsi\Lz} decays in each data set and in each \Lz category is determined by performing an extended unbinned maximum-likelihood fit to the $\jpsi\Lz$ mass distribution. 
The signal is parameterised by the sum of two Crystal Ball (CB)  functions~\cite{Skwarnicki:1986xj} combined with a Gaussian function. 
The two CB functions have a common peak position and width; one has a power-law tail on the lower side of the peak, the other on the upper side of the peak. 
The Gaussian function shares the same peak position as the two CB functions. 
The tail parameters and the relative fractions of the three signal components are fixed, for each data set, from fits to simulated \decay{\Lb}{\jpsi\Lz} decays. 
The combinatorial background is described by an exponential function. 
The background from \decay{\Bz}{\jpsi\KS} decays is described by a CB function with parameters fixed from simulated decays. 
Figure~\ref{fig:selection:mass} shows the $m(\jpsi\Lz)$ distribution and the result of the fits for each of the four data-taking years, with the two \Lz categories combined. 
The signal yields in the long and downstream categories of the 2011, 2012, 2015 and 2016 data are given in Table~\ref{tab:yields}.

\begin{table}[!tb]
    \caption{
    Signal yields in the long and downstream categories of the 2011, 2012, 2015 and 2016 data sets.
    }
    \centering
    \begin{tabular}{ccccc}
    \toprule
    & 2011 & 2012 & 2015 & 2016 \\
    \midrule
    Long & $1\,792\pm 46$ & $4\,099\pm 74$ & $\phantom{0}\,925\pm 34$ & $\phantom{0}6\,291\pm \phantom{0}88$ \\
    Downstream & $3\,030 \pm 59$ & $7\,904 \pm 96$ & $1\,722\pm 47$ &  $12\,809\pm 125$ \\
    \bottomrule
    \end{tabular}
    \label{tab:yields}
\end{table}

\section{Angular efficiency} 
\label{sec:Efficiency}

Both the detector acceptance and candidate selection affect the observed angular distribution of the candidates. 
As described in Ref.~\cite{LHCb-PAPER-2018-029}, the largest distortions of the angular distribution arise from kinematic requirements in the reconstruction and in the trigger. 
Corrections for the nonuniformity of the angular efficiency are determined using samples of simulated \decay{\Lb}{\jpsi\Lz} decays. 
The simulated samples are generated with isotropic decays of the \Lb baryon, the \Lz baryon and the \jpsi meson.
The resulting angular distribution is uniform in each of the five decay angles. 
After the selection procedure is applied, the angular distribution of the simulated decays is proportional to the full reconstruction and selection efficiency.
A full five-dimensional description is used to parameterise the angular distribution.
The parameterisation exploits the orthogonality of Legendre polynomials, $L_{j}(x)$, and of cosine functions. 
In its most general form, the distribution and hence the efficiency can be described by the sum
\begin{align}
    \varepsilon(\vec{\Omega}) = \sum_{rstuv} c_{rstuv} L_{r}(\cos\theta) L_{s}(\cos\theta_{l}) L_{t}(\cos\theta_{b}) L_{u}(\phi_{b}/\pi) \cos (v\phi_{l})~. 
    \label{eq:efficiency}
\end{align}
The coefficients $c_{rstuv}$ are determined by performing a moment analysis of the simulated sample. 

To describe the efficiency shape accurately, a large number of terms is needed in each dimension. 
An absolute normalisation of the efficiency is not needed in this analysis.
To reduce the complexity of the parameterisation, an iterative approach is used, where the efficiency model is constructed in stages. 
At the first stage, each dimension is parameterised independently and the simulated decays are corrected by the inverse of this simplified efficiency model. 
At the second stage, three-dimensional corrections are determined separately for ($\cos\theta_{l}$, $\phi_{l}$, $\cos\theta$) and for ($\cos\theta_{b}$, $\phi_{b}$, $\cos\theta$), which are subsequently applied to the simulated decays.
Finally, a five-dimensional correction is applied according to Eq.~\ref{eq:efficiency} with $r$, $s$, $t$, $u$ and $v$ between zero and two. 
Since the \mup and \mun from the \jpsi have almost identical interactions in the detector, the parameterisation is required to be symmetric in $\cos\theta_l$ and $\phi_l$ about zero such that only terms with even values of $s$ and $v$ are used in the efficiency model. 
This assumption is validated on simulated \decay{\Lb}{\jpsi\Lz} decays, generated with a more realistic decay model. 
A separate efficiency correction is derived for the long and downstream \Lz categories in each data-taking year.

\section{Angular moments} 
\label{sec:Moments}

The values of the angular terms normalised to the total rate, $M_{i} = J_{i}/(2J_1 + J_2)$, can be determined from the data by a moment analysis, 
\begin{align}
\begin{split}
    M_{i} &= \frac{1}{2 J_{1} + J_{2}} \int\limits_{\Omega} \frac{\deriv^{5}\Gamma}{\deriv\vec{\Omega}} g_{i}(\vec{\Omega}) \deriv\vec{\Omega}~,
\end{split}
\end{align}
through an appropriate choice of the functions $g_{i}(\vec{\Omega})$~\cite{Blake:2017une}.  
The integral can be estimated by a sum over the observed candidates, $c$, 
\begin{align}
M_{i} = \left(\sum\limits_{c=1}^{N} w_{c} g_{i}(\vec{\Omega}_c)\right)\Big/\left(\sum\limits_{c=1}^{N} w_{c}\right)~,
\end{align}
where the weights, $w_c$, are used to account for both background contamination and the non-uniform angular efficiency of the detector acceptance and the candidate selection and $N$ is the number of observed candidates. 
The background contamination is subtracted using the \sPlot technique~\cite{Pivk:2004ty} with $m({\jpsi\Lz})$ as a discriminating variable.

The analysis procedure is validated on \decay{\Bz}{\jpsi\KS} decays, where the \KS meson  subsequently decays to $\pip\pim$.
This decay has a similar topology to that of the \decay{\Lb}{\jpsi\Lz} decay but has an angular dependence that is uniform in $\cos\theta$, $\cos\theta_b$, $\phi_l$ and $\phi_b$ and depends only on $\sin^{2}\theta_l$, resulting in $M_1=\tfrac{1}{2}$ and the remaining moments being zero. 
The \decay{\Bz}{\jpsi\KS} candidates are selected in data in an analogous way to the \decay{\Lb}{\jpsi\Lz} candidates. 
The measured moments for the \decay{\Bz}{\jpsi\KS} decay are consistent with expectation and a $\chi^2$ comparison of the moments with their expected values yields a $p$-value of 12\%. 

The values of the moments for the \decay{\Lb}{\jpsi\Lz} decay at the three different centre-of-mass energies are given in Table~\ref{tab:moments}. 
The results from the long and downstream categories are compatible and are combined in the table.  
Systematic uncertainties on the moments are discussed in Section~\ref{sec:Systematics}. 
The values of moments $M_{11}$ to $M_{34}$ are consistent with zero, indicating a small production polarisation. 
The statistical covariance matrices for the moments are determined by bootstrapping the data set (\cf\ Ref.~\cite{efron:1979}) and repeating the analysis procedure. 
The correlation matrices for the moments are provided in Appendix~\ref{sec:appendix:correlation}. 
Figure~\ref{fig:angles} shows the background-subtracted angular projections of the five decay angles for the selected candidates.
Good agreement is seen between the data and the result of the moment analysis.
The values of the moments are also found to be in good agreement between \Lb and \Lbbar baryons, indicating that there is no significant difference in the production polarisation or decays of the \Lb and \Lbbar baryons. 
The numerical values of all moments and the corresponding covariance matrices are available as supplementary material to this article.

\begin{table}[!htb]
    \caption{
    Values of the 20 moments, $M_{i}$, measured in the data collected at 7, 8 and 13\tev centre-of-mass energies. 
    The long and downstream categories have been combined. 
    The first and second uncertainties are statistical and systematic, respectively. 
    }
    \centering
    \begin{tabular}{c|c|c|c}
       \toprule
        & 7\tev & 8\tev & 13\tev \\
        \midrule
        $M_{1}$ & $\phantom{+}0.374 \pm 0.007 \pm 0.003$ & $\phantom{+}0.373 \pm 0.004 \pm 0.002$ & $\phantom{+}0.380 \pm 0.003 \pm 0.001$ \\
        $M_{2}$ & $\phantom{+}0.253 \pm 0.014 \pm 0.005$ & $\phantom{+}0.254 \pm 0.008 \pm 0.003$ & $\phantom{+}0.239 \pm 0.006 \pm 0.002$ \\
        $M_{4}$ & $-0.286 \pm 0.017 \pm 0.008$ & $-0.268 \pm 0.011 \pm 0.009$ & $-0.273 \pm 0.008 \pm 0.006$ \\
        $M_{5}$ & $-0.157 \pm 0.025 \pm 0.008$ & $-0.181 \pm 0.015 \pm 0.007$ & $-0.179 \pm 0.011 \pm 0.005$ \\
        $M_{7}$ & $\phantom{+}0.051 \pm 0.029 \pm 0.005$ & $\phantom{+}0.025 \pm 0.018 \pm 0.003$ & $\phantom{+}0.022 \pm 0.013 \pm 0.002$ \\
        $M_{9}$ & $-0.017 \pm 0.029 \pm 0.005$ & $-0.011 \pm 0.018 \pm 0.003$ & $-0.027 \pm 0.013 \pm 0.002$ \\
        $M_{11}$ & $\phantom{+}0.005 \pm 0.014 \pm 0.004$ & $\phantom{+}0.003 \pm 0.009 \pm 0.004$ & $-0.005 \pm 0.006 \pm 0.002$ \\
        $M_{12}$ & $-0.004 \pm 0.018 \pm 0.005$ & $\phantom{+}0.010 \pm 0.011 \pm 0.004$ & $\phantom{+}0.006 \pm 0.008 \pm 0.003$ \\
        $M_{14}$ & $\phantom{+}0.007 \pm 0.025 \pm 0.007$ & $-0.015 \pm 0.016 \pm 0.007$ & $-0.009 \pm 0.012 \pm 0.003$ \\
        $M_{15}$ & $-0.027 \pm 0.032 \pm 0.008$ & $\phantom{+}0.009 \pm 0.021 \pm 0.008$ & $-0.006 \pm 0.016 \pm 0.005$ \\
        $M_{17}$ & $\phantom{+}0.008 \pm 0.039 \pm 0.006$ & $-0.002 \pm 0.025 \pm 0.004$ & $\phantom{+}0.011 \pm 0.018 \pm 0.003$ \\
        $M_{19}$ & $-0.006 \pm 0.038 \pm 0.004$ & $-0.015 \pm 0.025 \pm 0.004$ & $-0.003 \pm 0.018 \pm 0.002$ \\
        $M_{21}$ & $-0.015 \pm 0.037 \pm 0.008$ & $\phantom{+}0.007 \pm 0.022 \pm 0.005$ & $-0.032 \pm 0.016 \pm 0.005$ \\
        $M_{23}$ & $-0.001 \pm 0.028 \pm 0.007$ & $-0.022 \pm 0.017 \pm 0.003$ & $\phantom{+}0.018 \pm 0.012 \pm 0.002$ \\
        $M_{25}$ & $-0.029 \pm 0.064 \pm 0.010$ & $-0.001 \pm 0.038 \pm 0.008$ & $\phantom{+}0.044 \pm 0.029 \pm 0.006$ \\
        $M_{27}$ & $\phantom{+}0.059 \pm 0.051 \pm 0.007$ & $\phantom{+}0.014 \pm 0.030 \pm 0.005$ & $\phantom{+}0.038 \pm 0.023 \pm 0.006$ \\
        $M_{30}$ & $-0.000 \pm 0.023 \pm 0.004$ & $-0.028 \pm 0.014 \pm 0.005$ & $\phantom{+}0.008 \pm 0.010 \pm 0.003$ \\
        $M_{32}$ & $-0.001 \pm 0.021 \pm 0.005$ & $\phantom{+}0.013 \pm 0.014 \pm 0.004$ & $-0.022 \pm 0.010 \pm 0.003$ \\
        $M_{33}$ & $\phantom{+}0.019 \pm 0.021 \pm 0.005$ & $-0.017 \pm 0.013 \pm 0.003$ & $-0.007 \pm 0.009 \pm 0.002$ \\
        $M_{34}$ & $\phantom{+}0.017 \pm 0.021 \pm 0.004$ & $\phantom{+}0.033 \pm 0.013 \pm 0.004$ & $\phantom{+}0.008 \pm 0.009 \pm 0.002$ \\
        \bottomrule
    \end{tabular}
    \label{tab:moments}
\end{table}

\begin{figure}
    \begin{center}
    \includegraphics[width=0.32\linewidth]{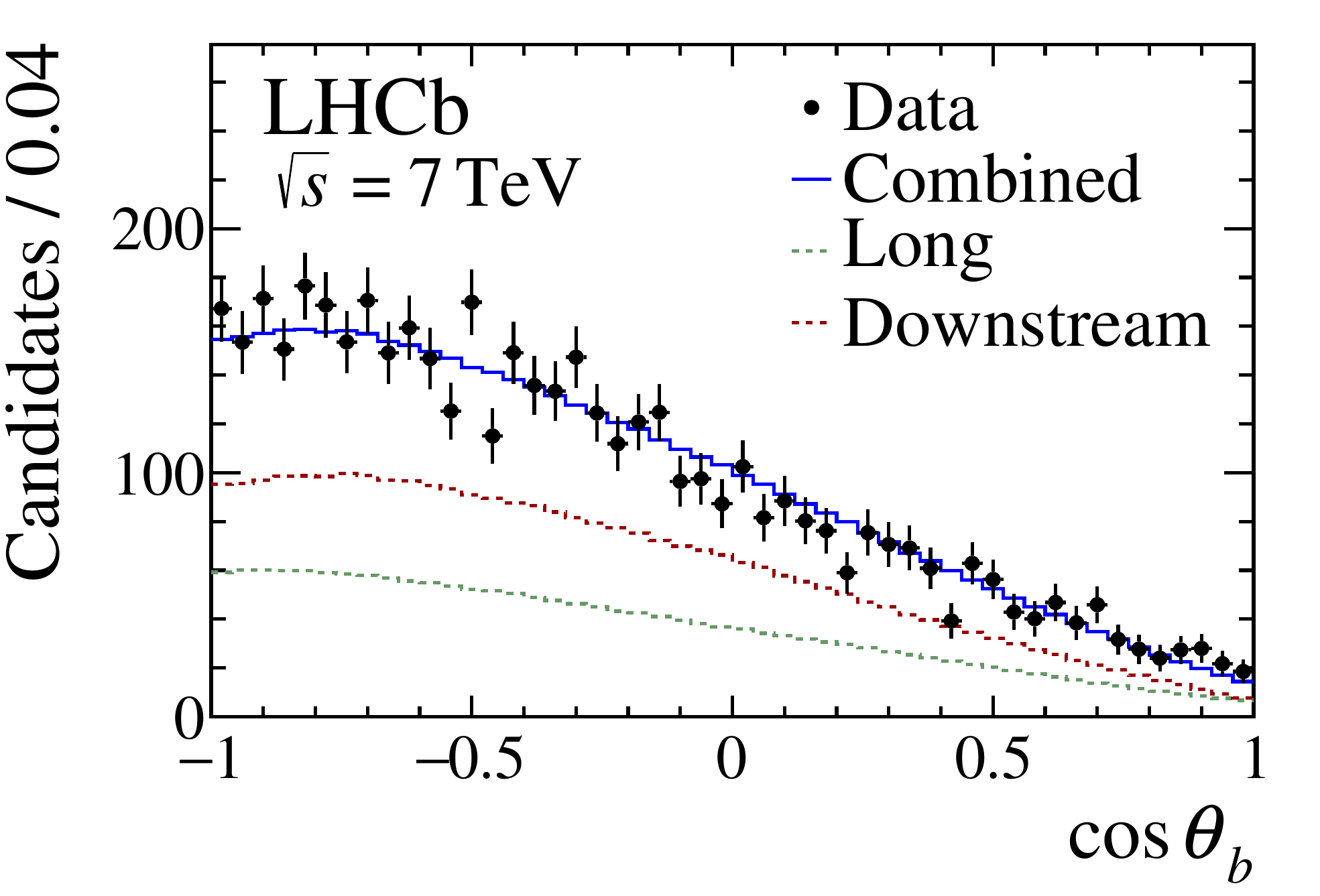} 
    \includegraphics[width=0.32\linewidth]{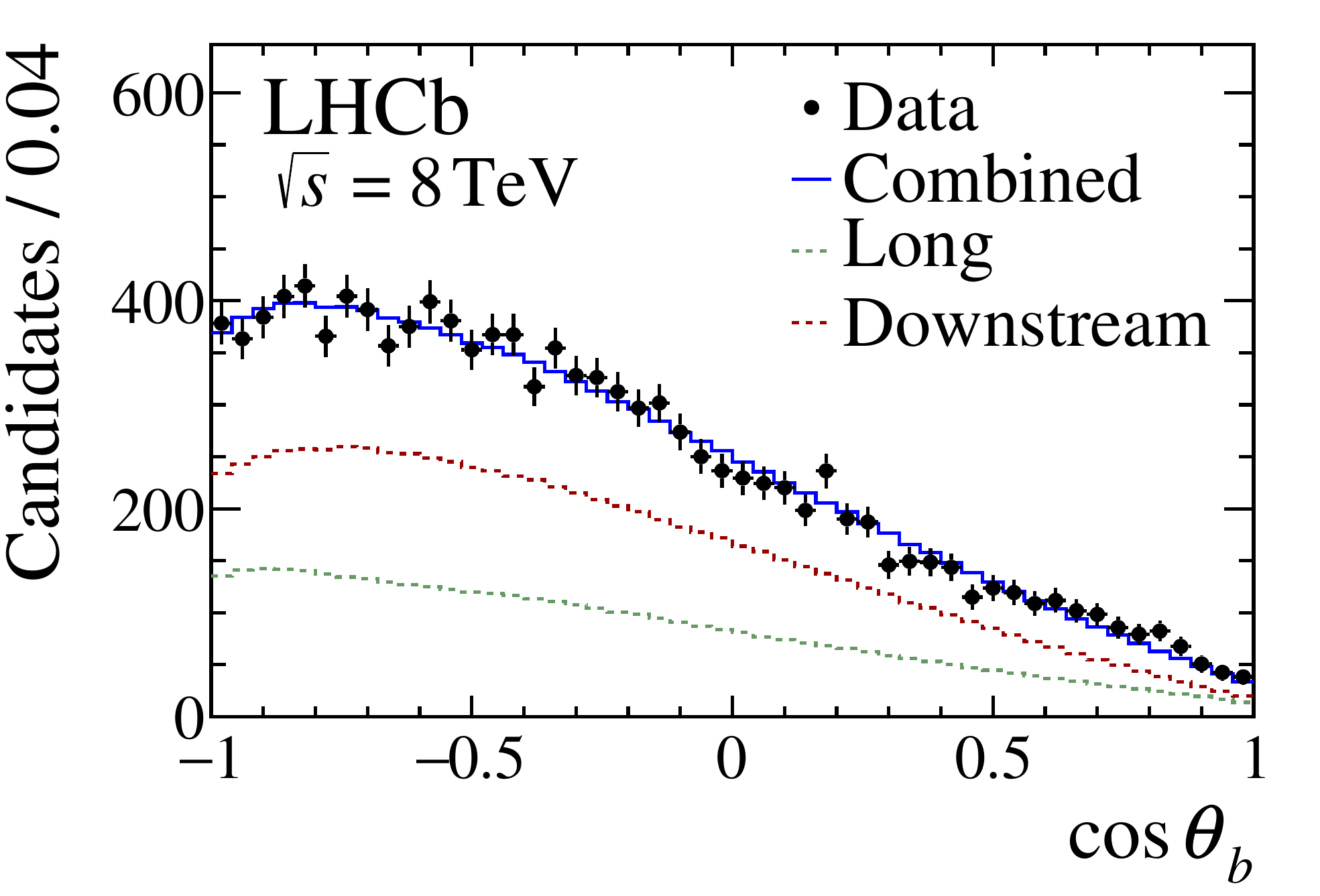} 
    \includegraphics[width=0.32\linewidth]{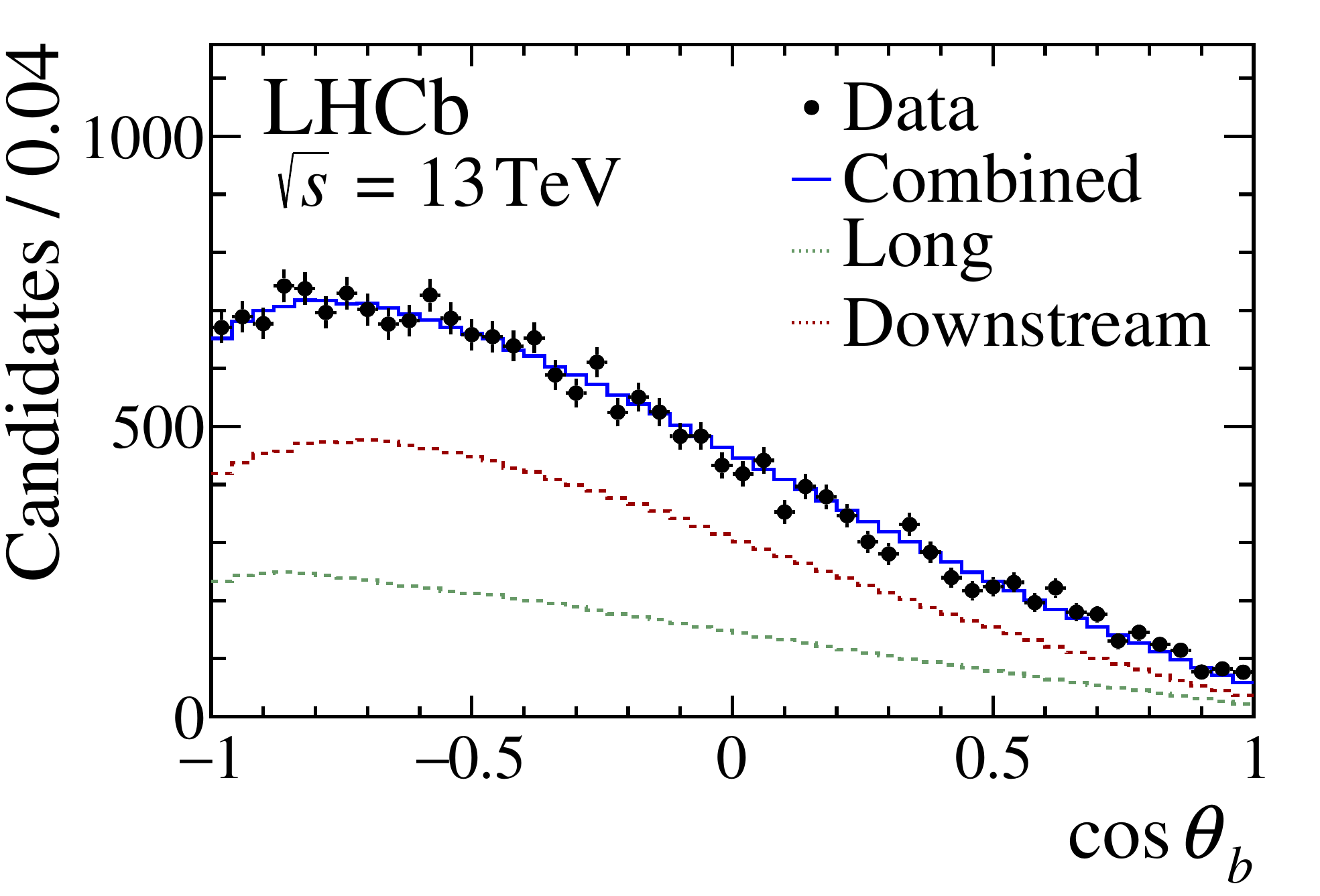} \\
    \includegraphics[width=0.32\linewidth]{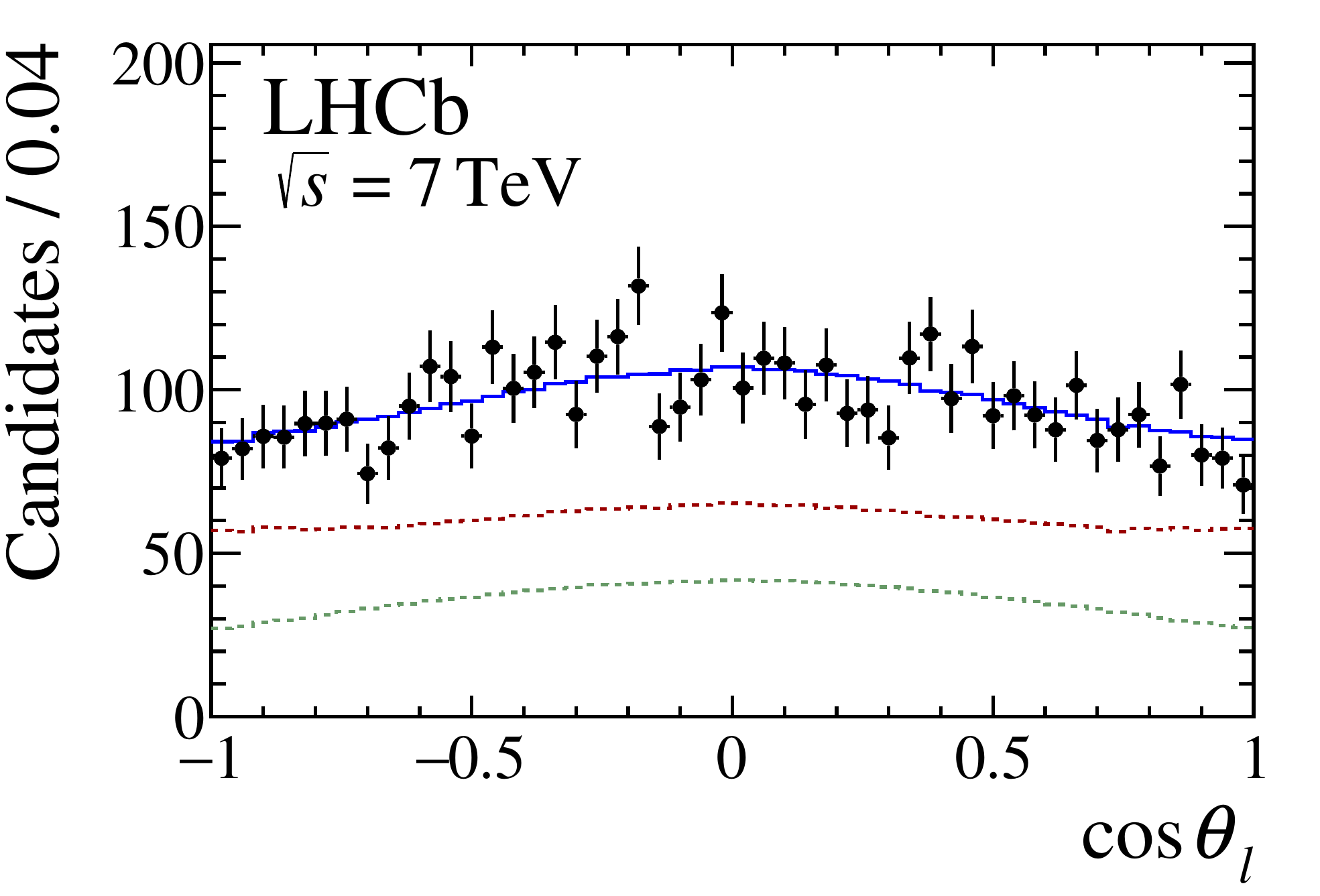} 
    \includegraphics[width=0.32\linewidth]{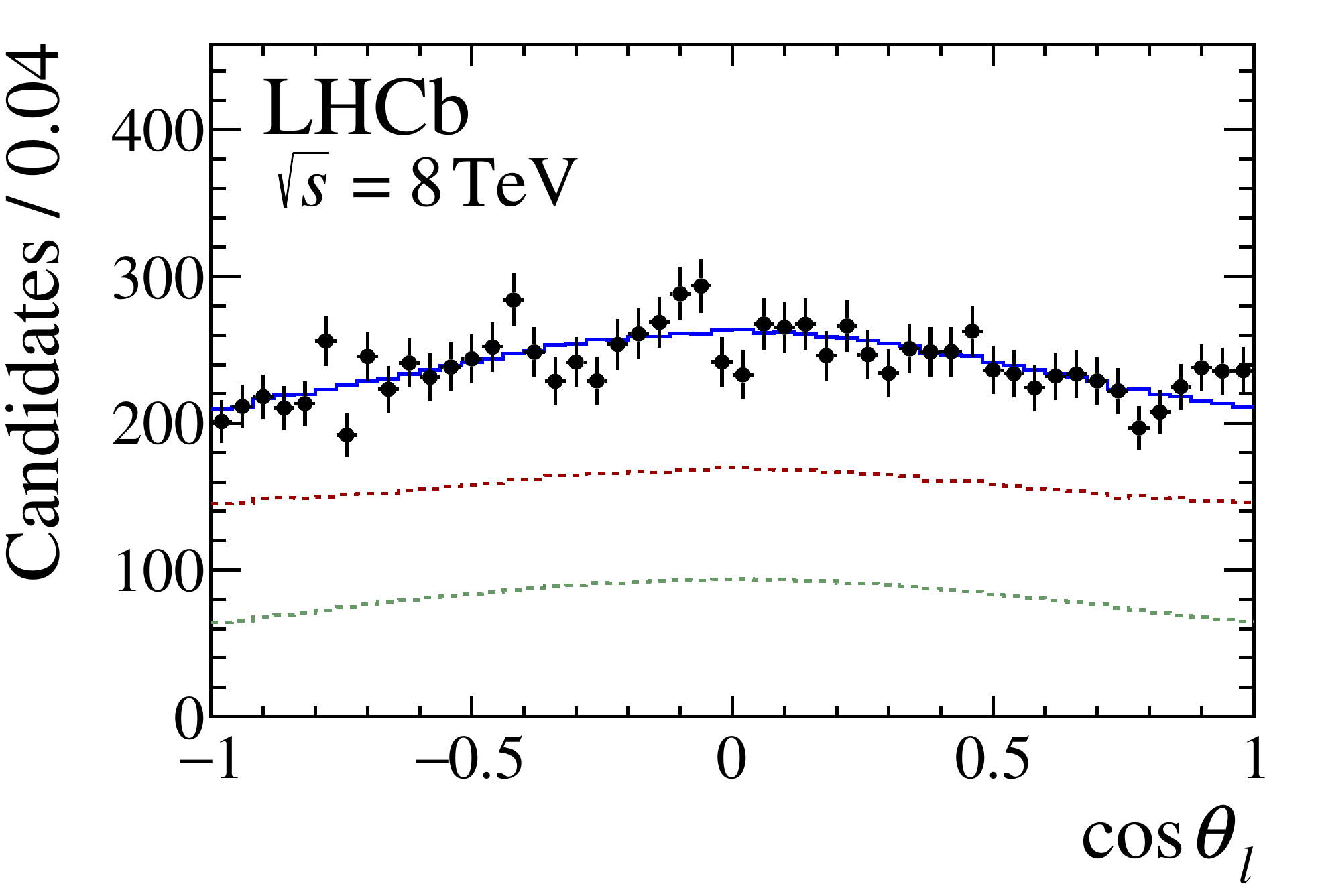} 
    \includegraphics[width=0.32\linewidth]{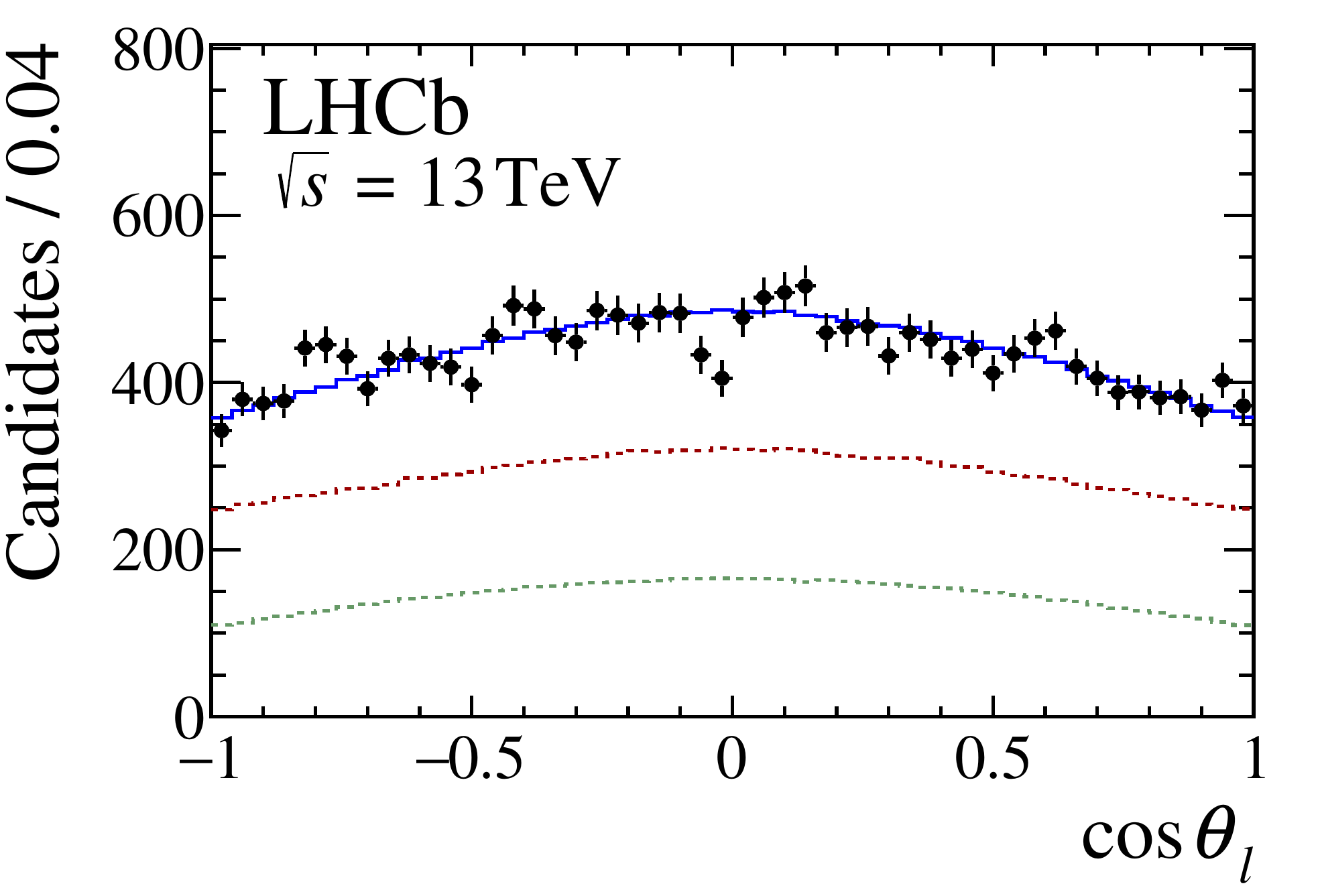} \\
    \includegraphics[width=0.32\linewidth]{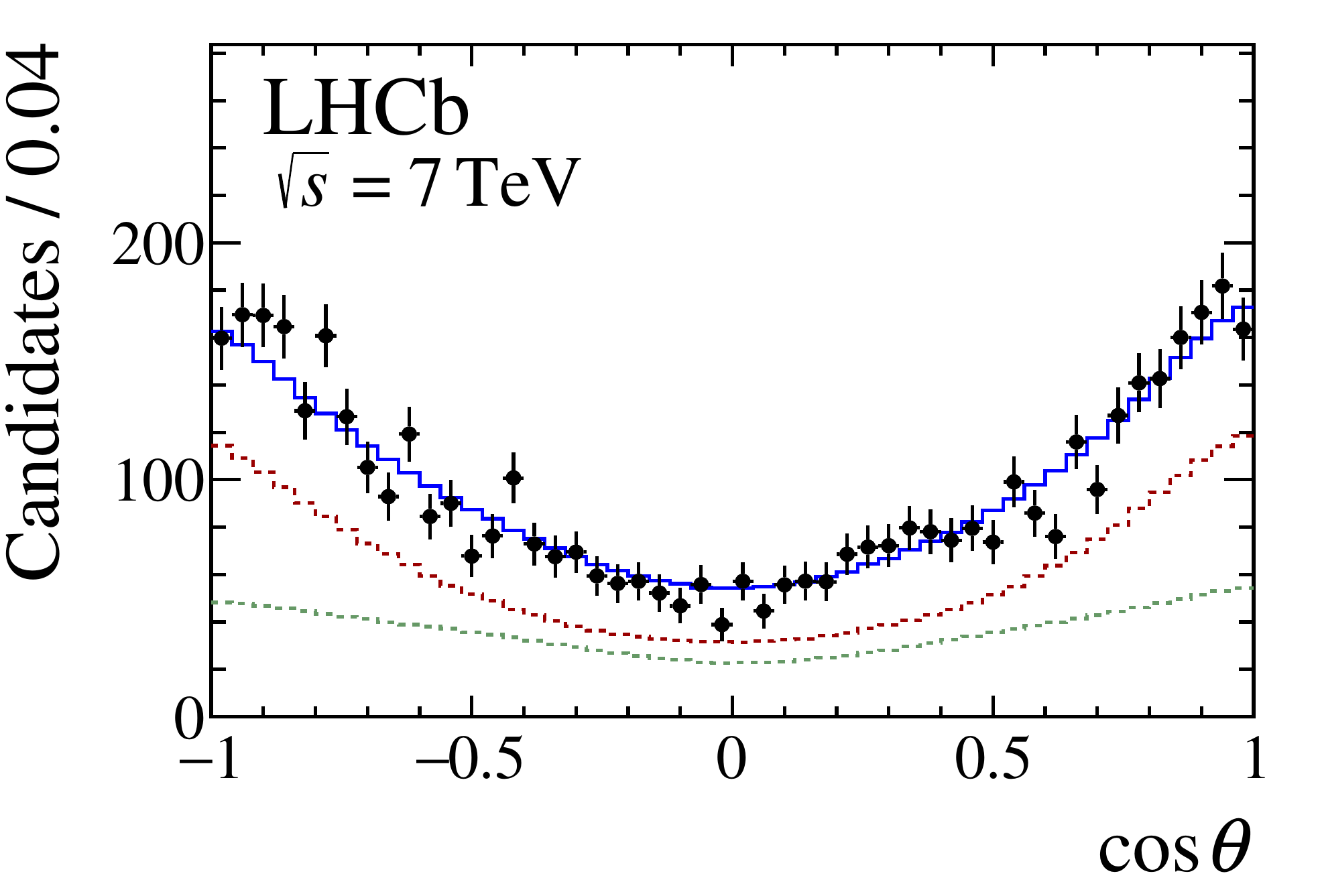} 
    \includegraphics[width=0.32\linewidth]{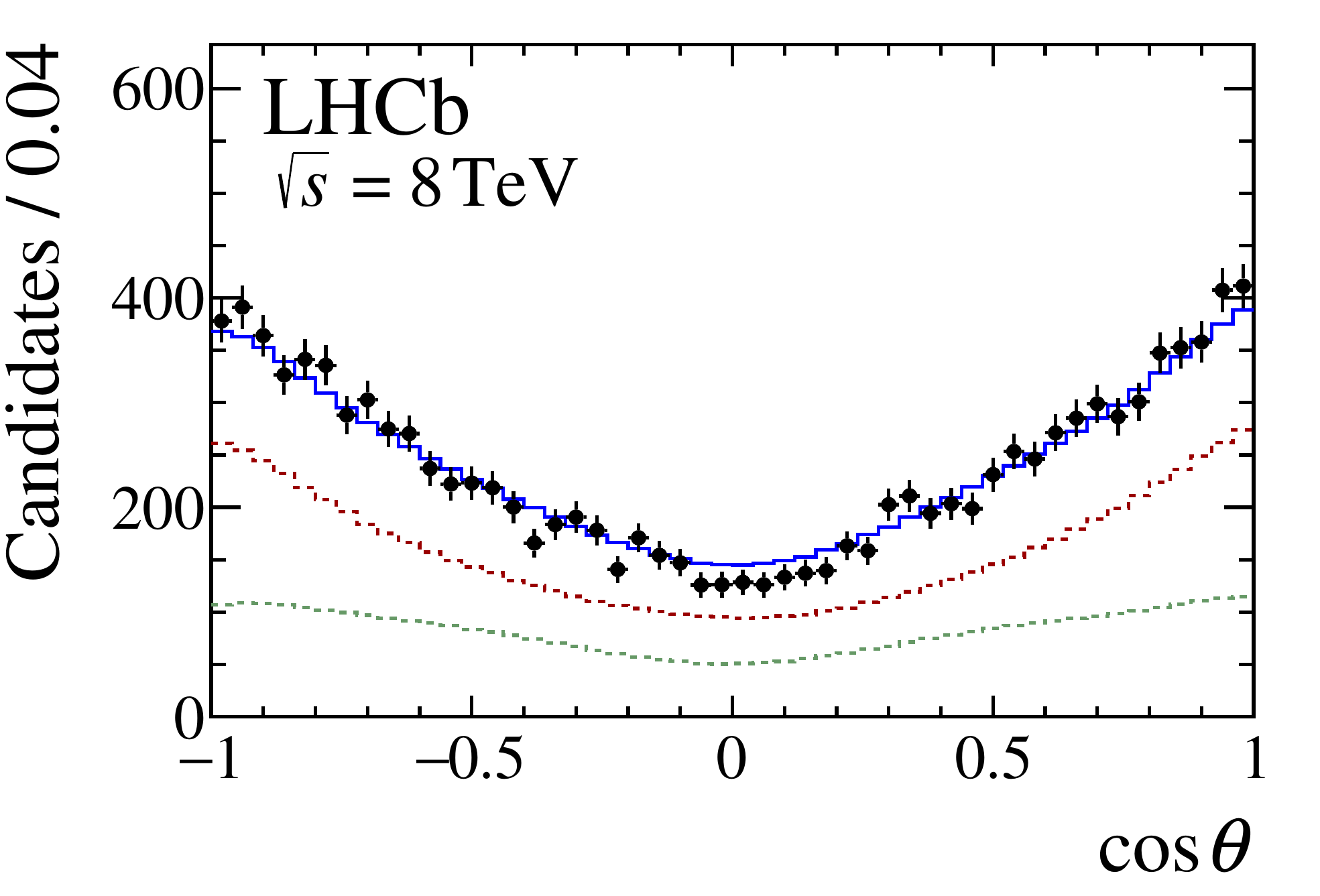} 
    \includegraphics[width=0.32\linewidth]{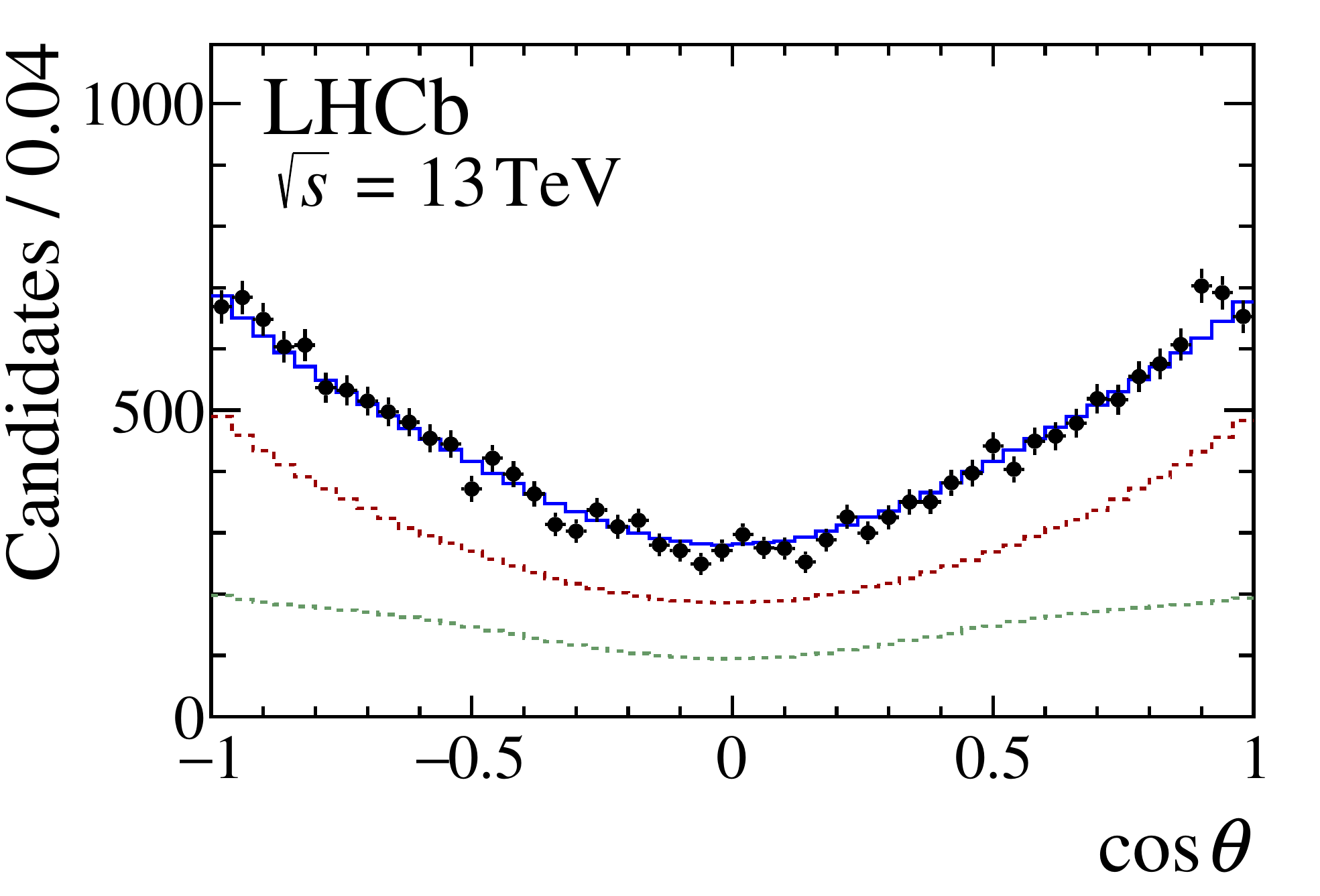} \\
    \includegraphics[width=0.32\linewidth]{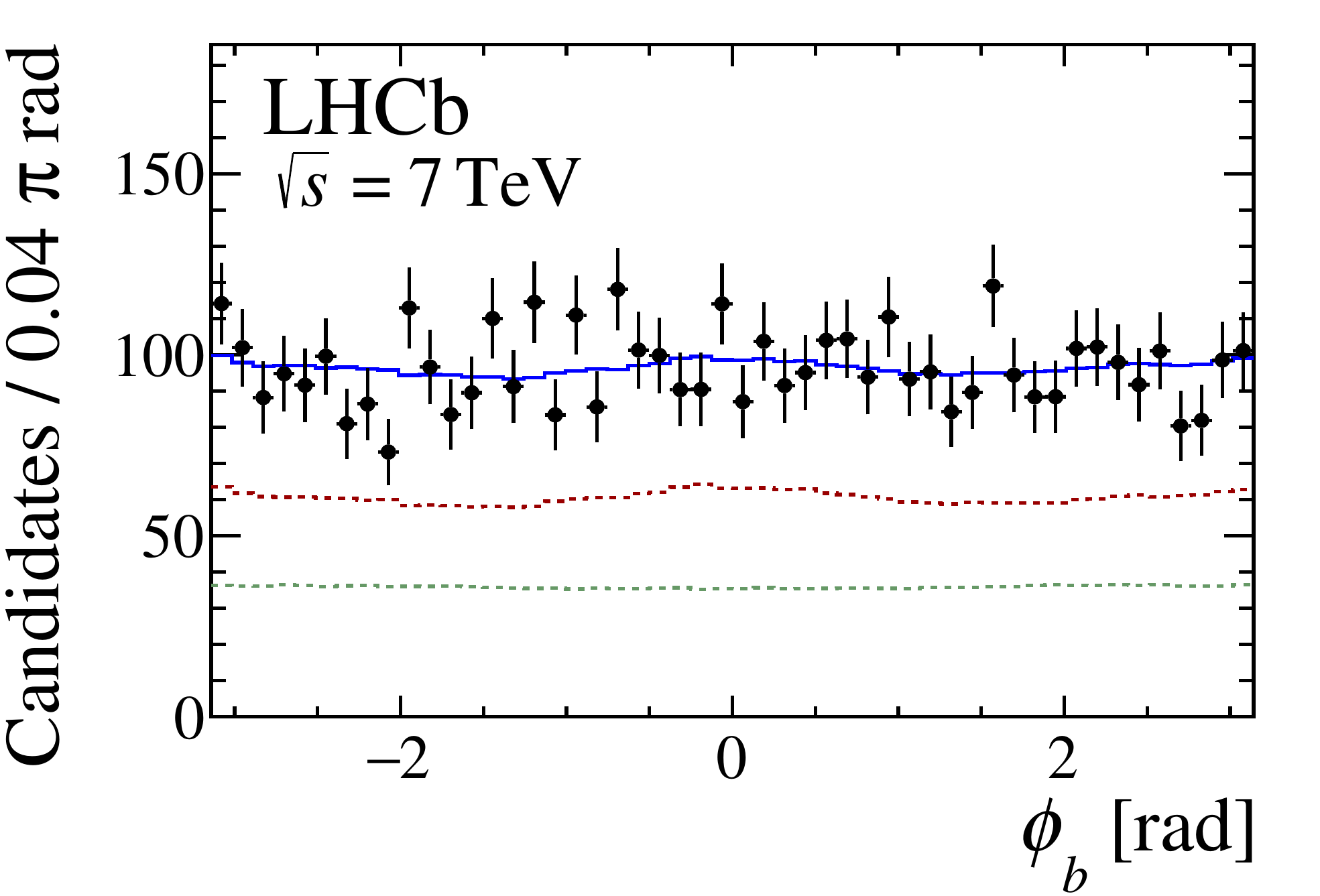} 
    \includegraphics[width=0.32\linewidth]{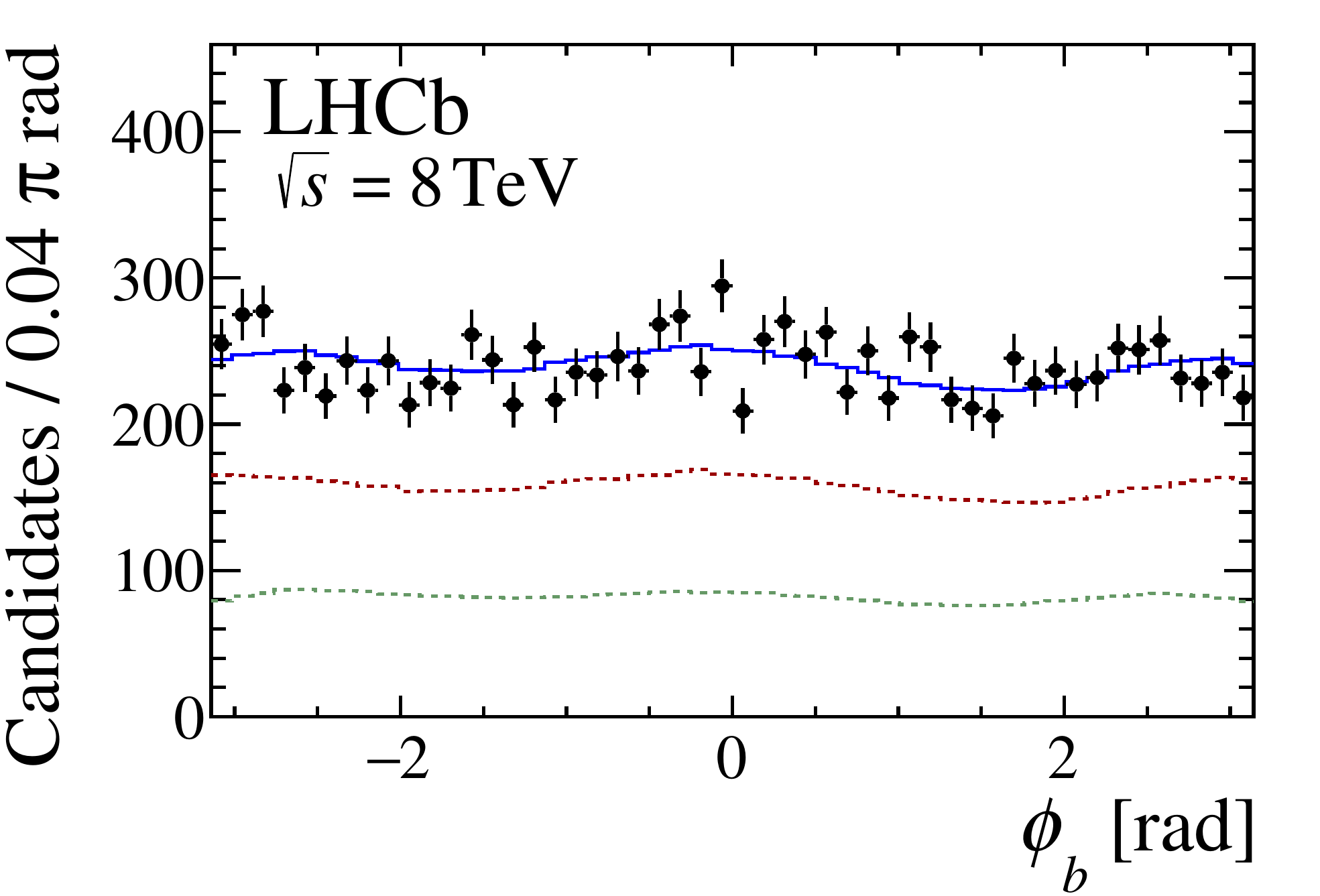} 
    \includegraphics[width=0.32\linewidth]{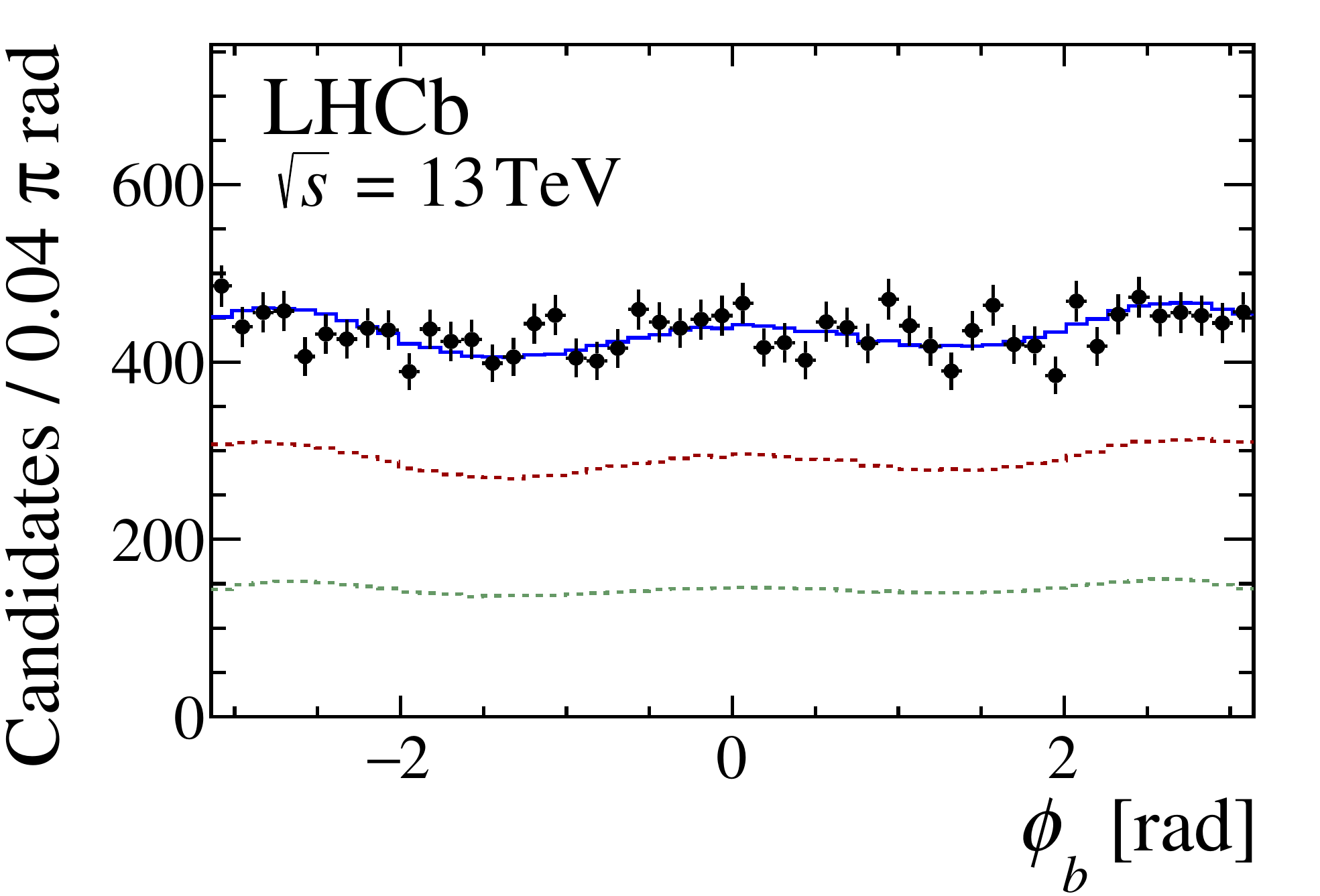} \\
    \includegraphics[width=0.32\linewidth]{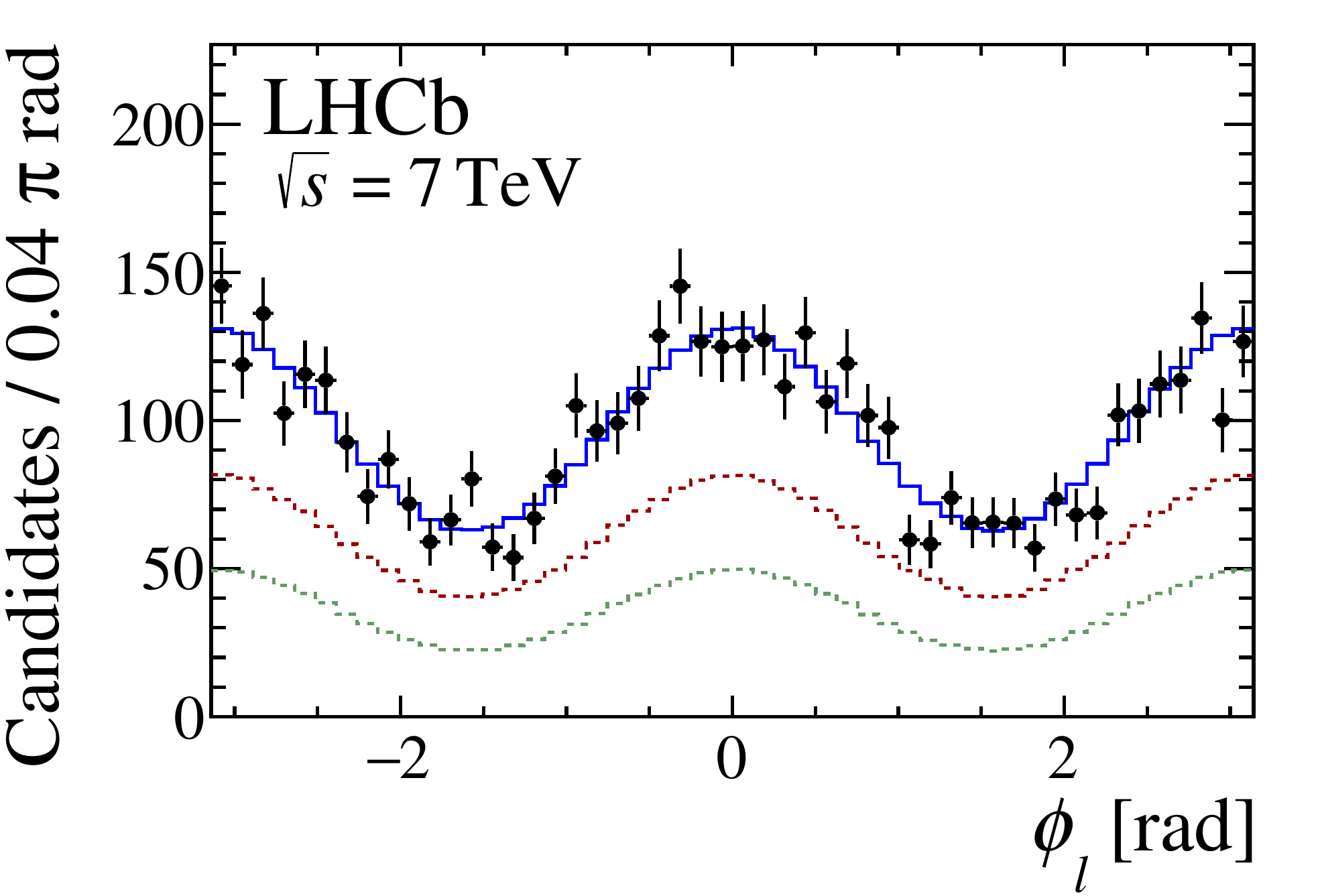} 
    \includegraphics[width=0.32\linewidth]{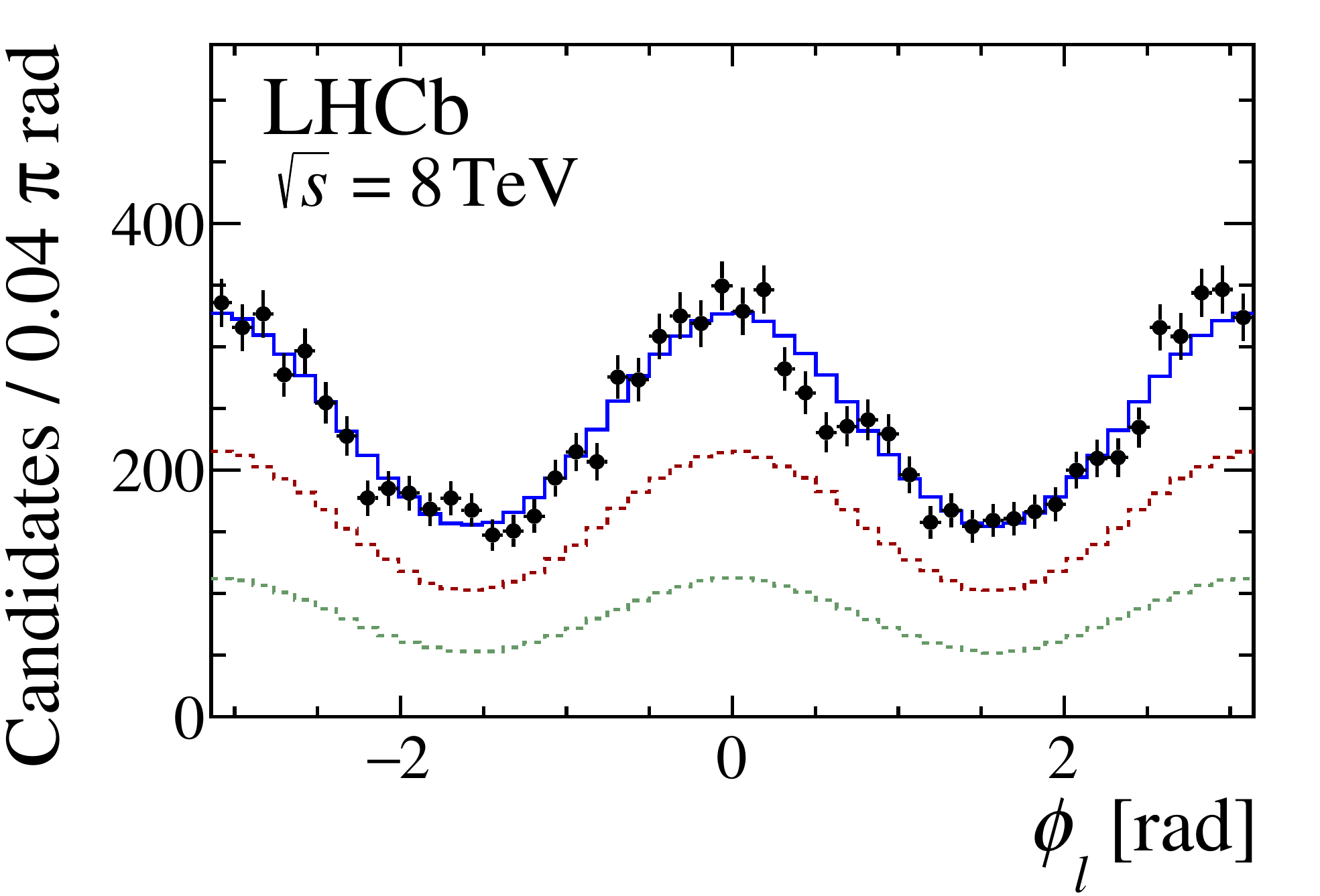} 
    \includegraphics[width=0.32\linewidth]{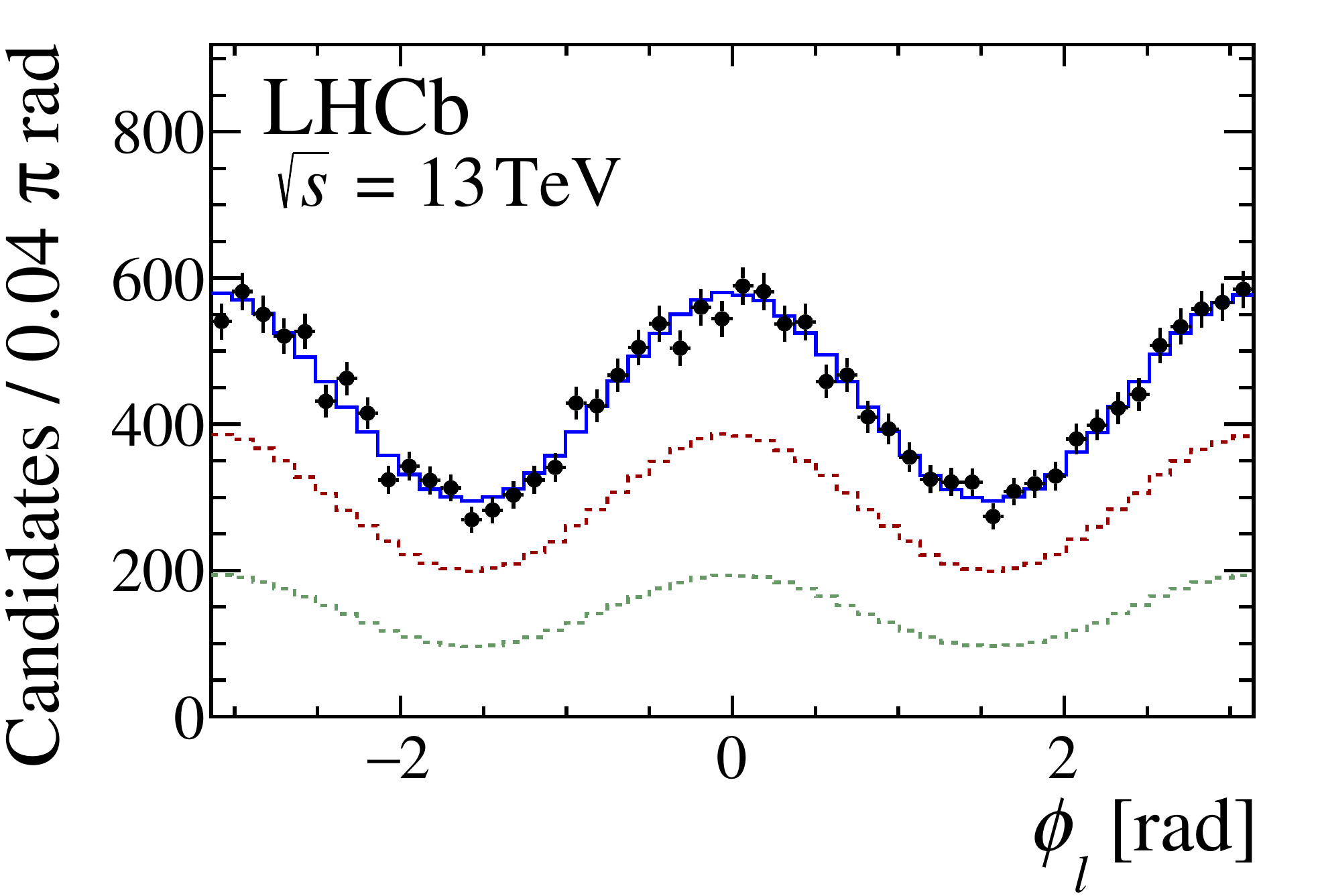}
    \end{center}
    \caption{
    Angular distributions of $\cos\theta_b$, $\cos\theta_l$, $\cos\theta$, $\phi_b$ and $\phi_l$ for the background-subtracted candidates. 
    The long and downstream categories for the different data-taking years have been combined. 
    The result of the moment analysis, folded with the angular efficiency, is overlaid. The contribution from the long and downstream categories are indicated by the green and red lines, respectively.
    }
    \label{fig:angles}
\end{figure}

\section{Systematic uncertainties}
\label{sec:Systematics}

Sources of systematic uncertainty are considered if they either impact the fit to the $m(\jpsi\Lz)$ distribution, and the subsequent background subtraction, or would directly bias the measured angular distribution. 
The various sources of systematic uncertainty on this measurement are discussed below and summarised in Table~\ref{tab:systematics}. 

A systematic uncertainty is assigned to cover the knowledge of the signal lineshape parameters by repeating the analysis 1000 times, varying the lineshape parameters within their uncertainties.
The resulting systematic uncertainty is given by the standard deviation of the moments evaluated with the different variations. 

The impact of statistical uncertainty on the efficiency model, due to the limited size of the simulated samples, is determined by bootstrapping the simulated samples 1000 times and rederiving the efficiency models.
For each bootstrap, a new set of efficiency coefficients, $c_{rstuv}$, is determined and the angular moments are reevaluated.
For each moment, the standard deviation of the distribution of the difference between the new and the nominal values is assigned as a systematic uncertainty. 

To evaluate the impact of the limited number of terms used for the efficiency model, a new parameterisation is determined that allows for higher-order terms in each dimension. 
Pseudoexperiments are then generated from the higher-order model and the values of the moments determined from each pseudoexperiment using the nominal model. The average bias on the determined value of the moments and its uncertainty are added in quadrature and are assigned as a systematic uncertainty. 

A systematic uncertainty is assigned to cover the choice of criteria used to match reconstructed and true particles in the simulation. 
This uncertainty is evaluated using pseudoexperiments that are generated from an efficiency model derived with a less strict set of matching requirements. The moments are then evaluated with the nominal model. As before, the average bias on the determined value of the moments and its uncertainty are added in quadrature and are assigned as a systematic uncertainty. 

The impact of neglecting the detector's angular resolution in the analysis is explored using pseudoexperiments in which the simulated angles are smeared according to the resolution. 
The resolution, determined using simulated decays, is approximately $3\mrad$ in $\theta$ and $\theta_l$, $20\mrad$ in $\theta_b$, $10\mrad$ in $\phi_l$ and $45\mrad$ in $\phi_b$. 
The resolution of the long and downstream categories are similar after constraining the masses of the \jpsi and \Lz candidates to their known values. 
The angular moments are then determined from the pseudoexperiments, neglecting the resolution. 
The average bias on the determined value of the moments and its uncertainty are added in quadrature and are assigned as a systematic uncertainty. 
The analysis procedure also assumes that the mass and angular variables factorise for both the signal and the background. 
No significant correlation is found between the mass and angular distribution of simulated \decay{\Lb}{\jpsi\Lz} decays. 
The variables are also found to be uncorrelated for the combinatorial background. 
However, a correlation is seen between the mass and angular distributions of misidentified \decay{\Bz}{\jpsi\KS} decays. 
The impact of neglecting this correlation is tested using pseudoexperiments, with the mass and angular distributions of the \decay{\Bz}{\jpsi\KS} decays taken from a detailed simulation. 
The values of the moments are then determined neglecting the correlation and the resulting bias is taken as a systematic uncertainty. 
In principle there is also an effect arising from neglecting the precession of the \Lz-baryon spin in the external magnetic field of the experiment.
The precession is small due to the small size of the integrated field between the production and decay points of the \Lz baryon.

The track-reconstruction and muon-identification efficiency of the LHCb detector are determined from data, in bins of \pt and $\eta$, using a tag-and-probe approach with \decay{\jpsi}{\mumu} decays~\cite{LHCb-DP-2014-002}. 
The resulting corrections to the simulation are small and are neglected in the analysis. 
The impact of neglecting these corrections is evaluated using pseudoexperiments. 
The pseudoexperiments are generated from an efficiency model that takes into account the corrections. 
The moments are then determined using a model that neglects the corrections and a systematic uncertainty is assigned based on the average bias on the moments and its uncertainty. 

The trigger efficiency of the hardware trigger is also determined in data, as a function of the muon \pt, using the method described in Ref.~\cite{LHCb-DP-2012-004}. 
The impact of the resulting corrections to the simulation is again investigated with pseudoexperiments. 
The pseudoexperiments are generated taking into account corrections to the trigger efficiency and the moments are determined neglecting the corrections. 
The resulting uncertainty is assigned based on the average bias and its uncertainty. 

A systematic uncertainty is assigned to the kinematic weighting of the simulated samples using pseudoexperiments. The pseudoexperiments are generated using the nominal model from which moments are determined using an efficiency model that neglects the kinematic corrections.
Again, the average bias and its uncertainty are added in quadrature and are assigned as the systematic uncertainty. 

Finally, a systematic uncertainty is evaluated to cover the uncertainty on the beam crossing angle at the LHCb interaction point. 
This is estimated using simulated events in which the crossing angle is varied. 
The resulting systematic uncertainty is negligible. 

The total systematic uncertainty on each moment is determined by summing the individual sources of uncertainty in quadrature. The resulting values are given in Table~\ref{tab:moments}. 
The systematic uncertainty is typically less than half the size of the statistical uncertainty on a given moment. 
Correlated systematic uncertainties between different moments are found to be small as are correlations between the different data sets. 
Correlations between systematic uncertainties are therefore neglected when determining the decay amplitudes and production polarisation. 

\begin{table}[!tb]
    \caption{
    Systematic uncertainties on the angular moments. 
    The largest value amongst the moments is given for each source. 
    The total systematic uncertainty varies from 0.002 to 0.010, depending on the moment considered.
    The sources are described in the text. 
    }
    \centering
    \begin{tabular}{lc}
    \toprule
    Source & Uncertainty\\
    \midrule
    Mass model & $0.003$ \\
    Simulation sample size & $0.006$ \\
    Polynomial order &  $0.004$ \\
    Truth matching criteria & $0.007$ \\
    Angular resolution & $0.003$ \\
    Factorisation of mass and angles & $0.003$ \\
    Tracking and muon-identification efficiency & $0.005$ \\
    Trigger efficiency modelling & $0.003$ \\
    Kinematic weighting & $0.006$ \\
    Beam-crossing angle & $0.001$ \\
    \bottomrule
    \end{tabular}
    \label{tab:systematics}
\end{table}

\section{Decay amplitudes and production polarisation} 
\label{sec:Results}

The decay amplitudes and the production polarisation are determined from the moments using a Bayesian analysis. The marginalisation over unwanted parameters is performed using Markov Chain Monte Carlo, with the Metropolis-Hastings algorithm employed to sample points in the parameter space~\cite{Metropolis1953,Hastings:1970aa}. 
The likelihood at each point in the parameter space is given by
\begin{align}
    L = \left[ \prod\limits_{{\rm data\,set}\,j} {\rm exp}( -\tfrac{1}{2} \vec{D}_{j}^{\rm T} C^{-1}_{j} \vec{D}_{j}^{}) \right] \times {\rm exp} \left( - \frac{1}{2} \left(\frac{ \alpha_{\Lz} - \alpha_{\Lz}^{\text{BES}}}{\sigma(\alpha^{\text{BES}}_{\Lz})} \right)^{2} \right)~,
\end{align}
where $\vec{D}_{j}$ is a vector representing the difference between the measured values of the moments and the values of the moments at that point in the parameter space and $C_{j}$ is the covariance matrix combining the statistical and systematic uncertainties on the moments. 
The last term in the likelihood originates from the external constraints from BES\,III on the value of $\alpha_{\Lz}$.
In this analysis, the recent BES\,III result~\cite{Ablikim:2018zay} for the \Lz asymmetry parameter is used. 
Averaging the BES\,III values for \Lz and \Lbar decays yields $\alpha_{\Lz}^{\text{BES}} = 0.754$ with an uncertainty $\sigma(\alpha^{\text{BES}}_{\Lz}) = 0.003$.
The value of $\alpha_{\Lz}$ and the values of the complex amplitudes $a_{\pm}$ and $b_{\pm}$ are shared between the different data sets but the polarisation is allowed to differ between different centre-of-mass energies. 
The Bayesian analysis procedure has been validated for both small and large values of the polarisation using pseudoexperiments.

\begin{figure}[!tb]
    \begin{center}
    \includegraphics[width=0.32\linewidth]{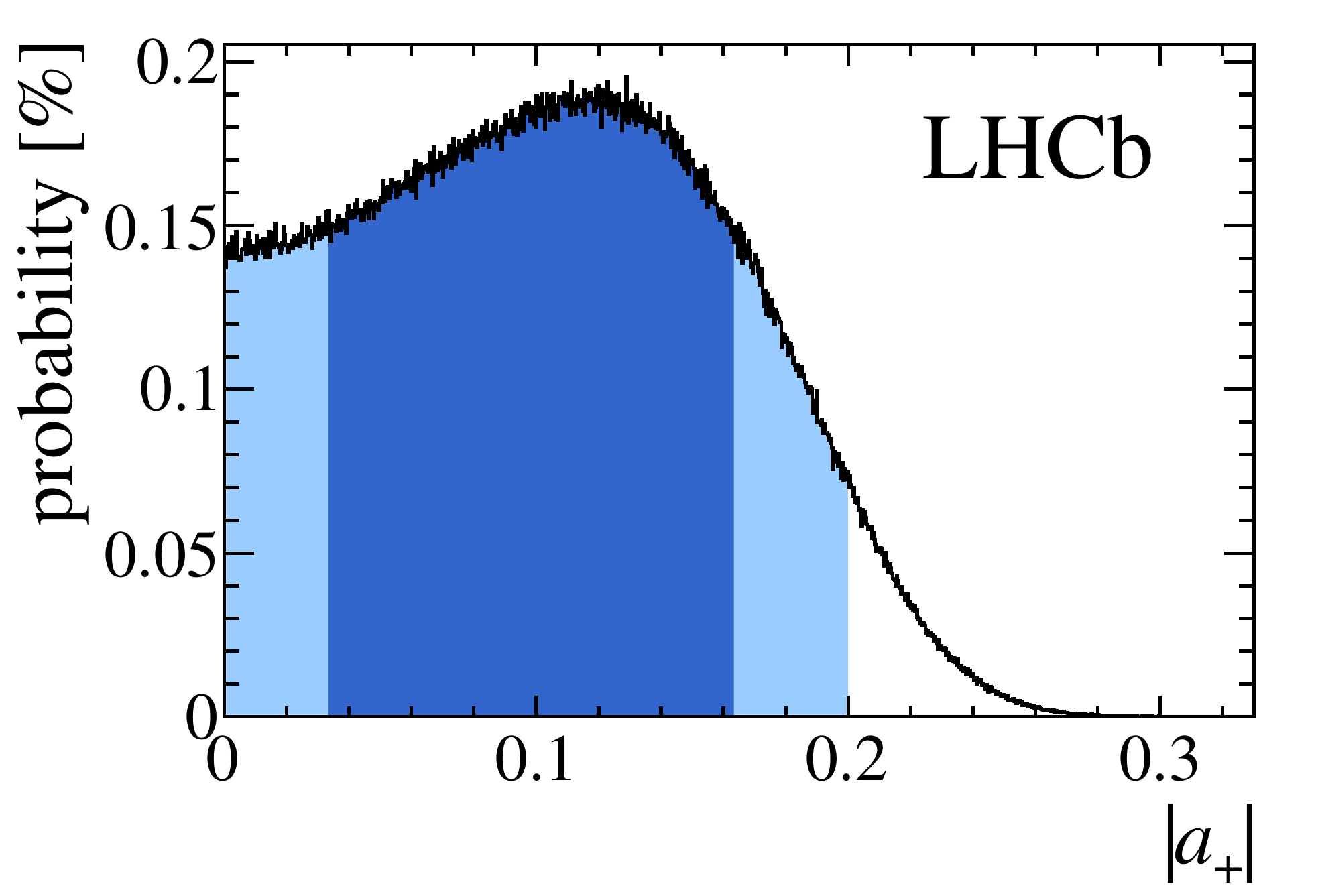} 
    \includegraphics[width=0.32\linewidth]{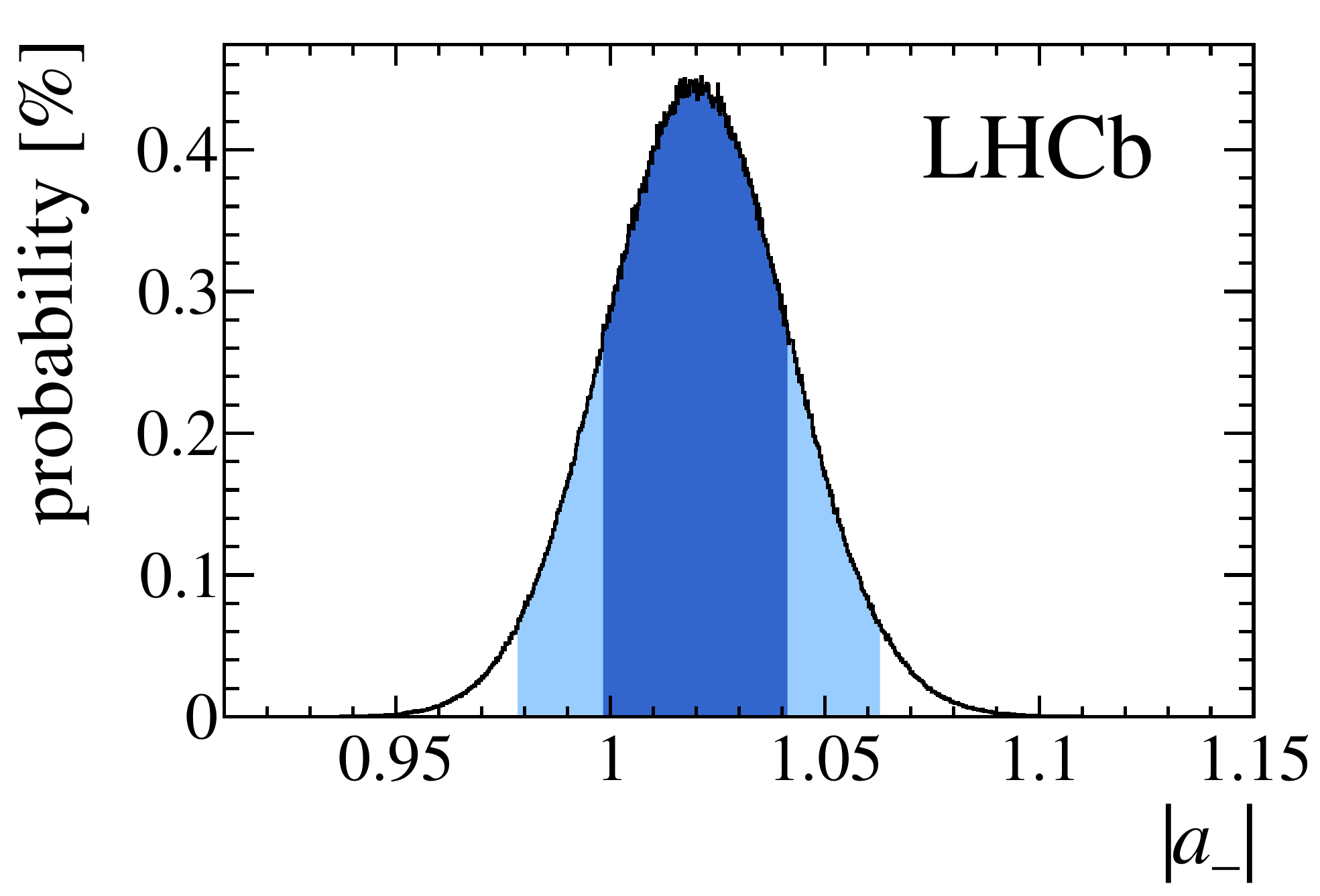} 
    \includegraphics[width=0.32\linewidth]{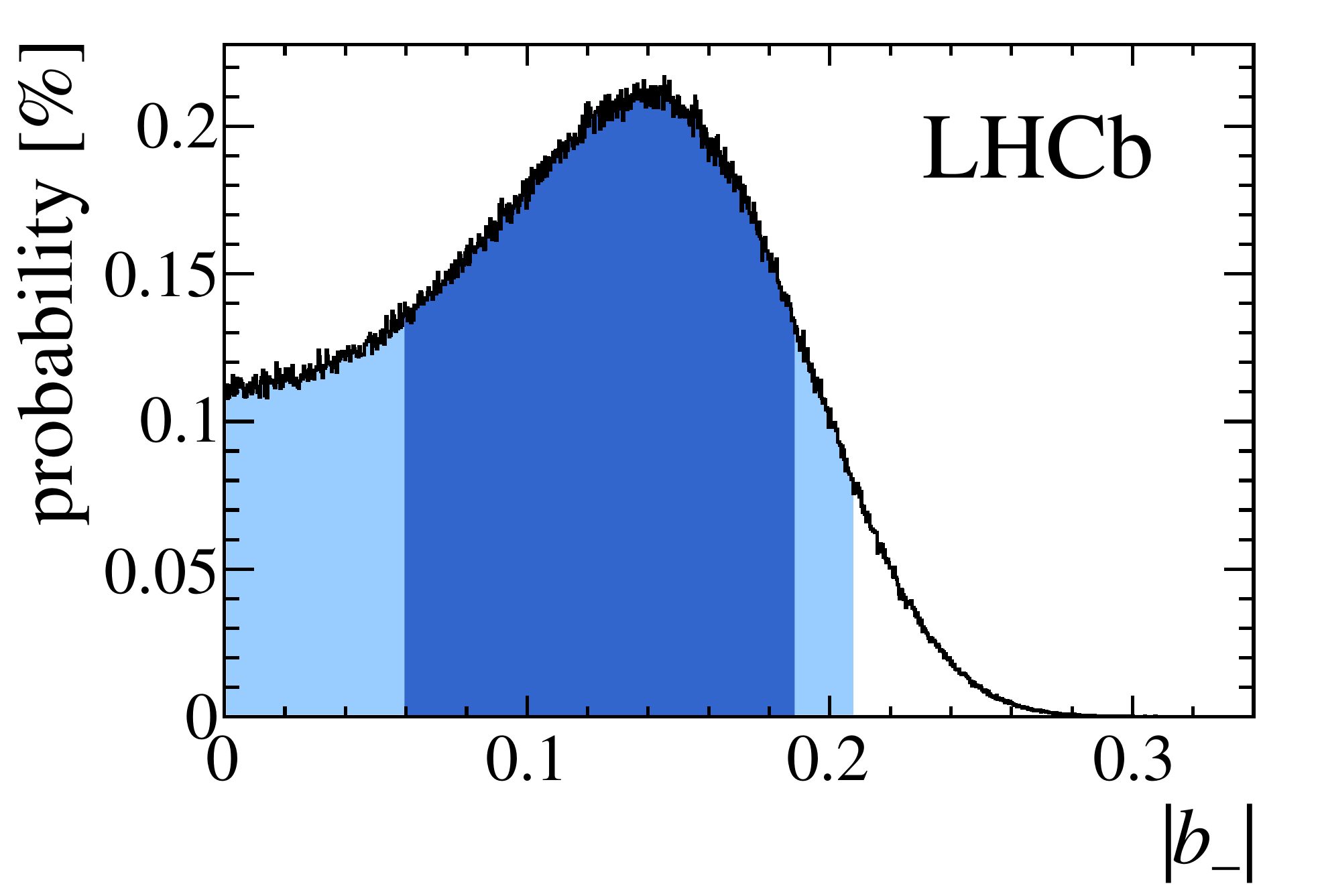} \\
    \includegraphics[width=0.32\linewidth]{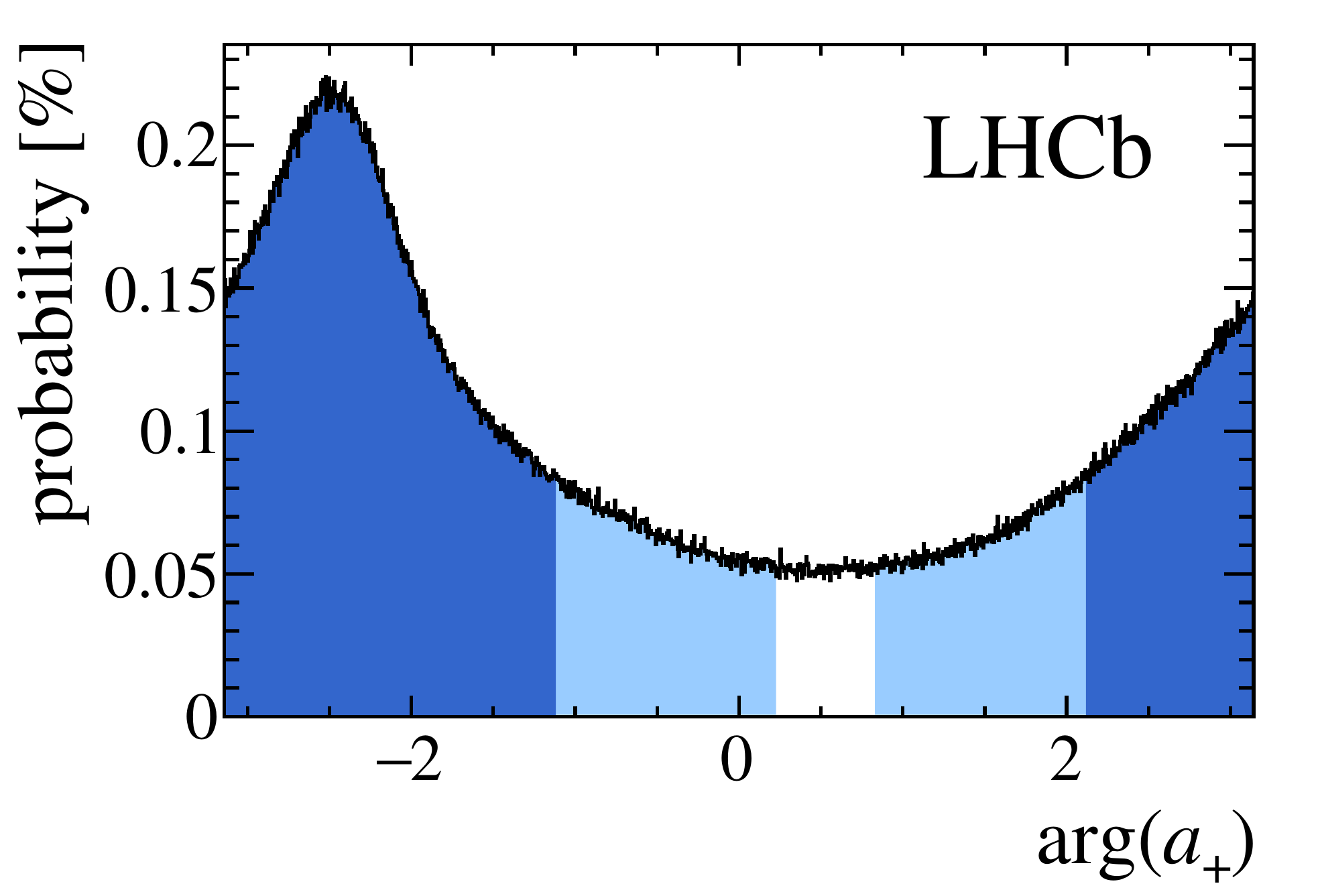} 
    \includegraphics[width=0.32\linewidth]{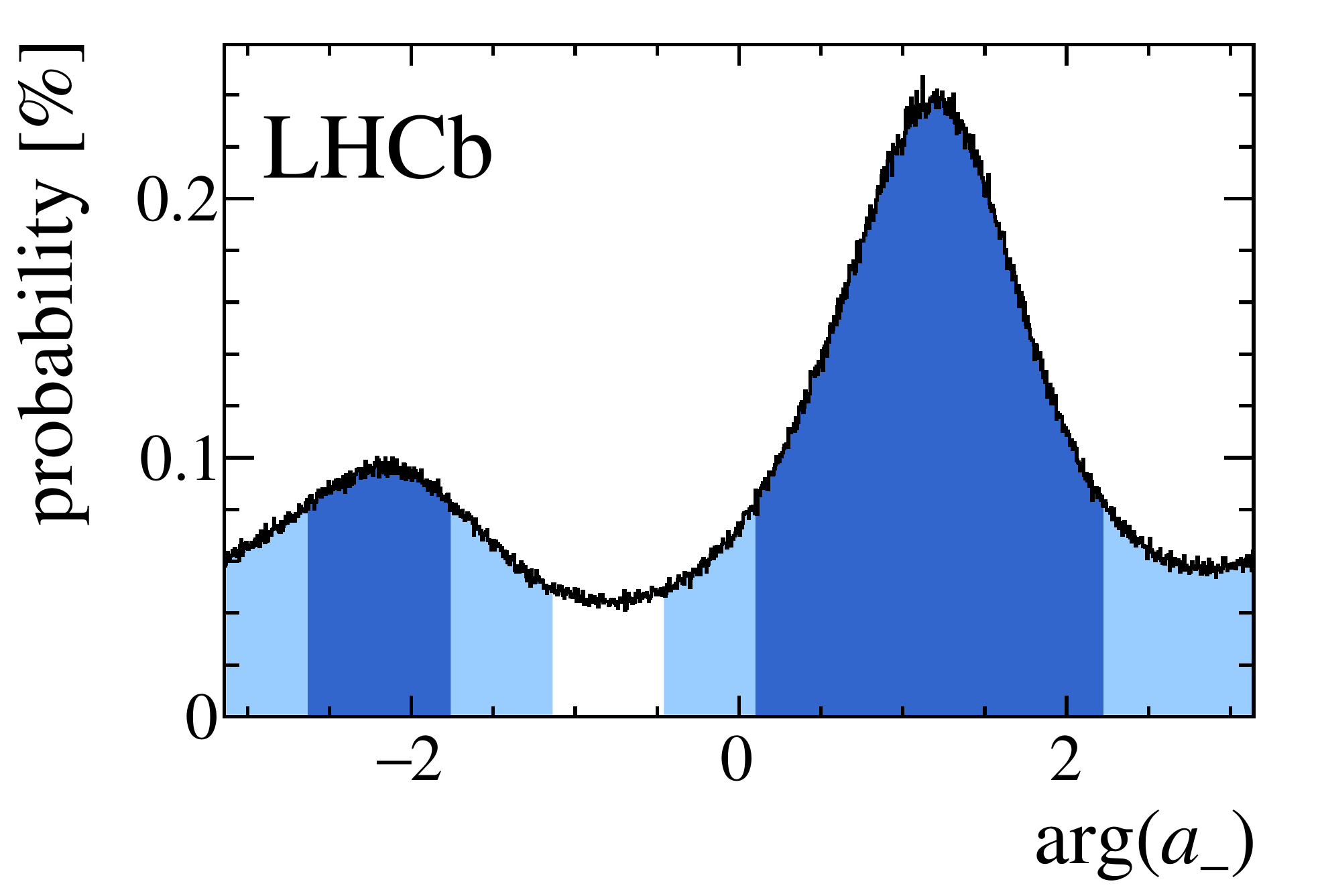} 
    \includegraphics[width=0.32\linewidth]{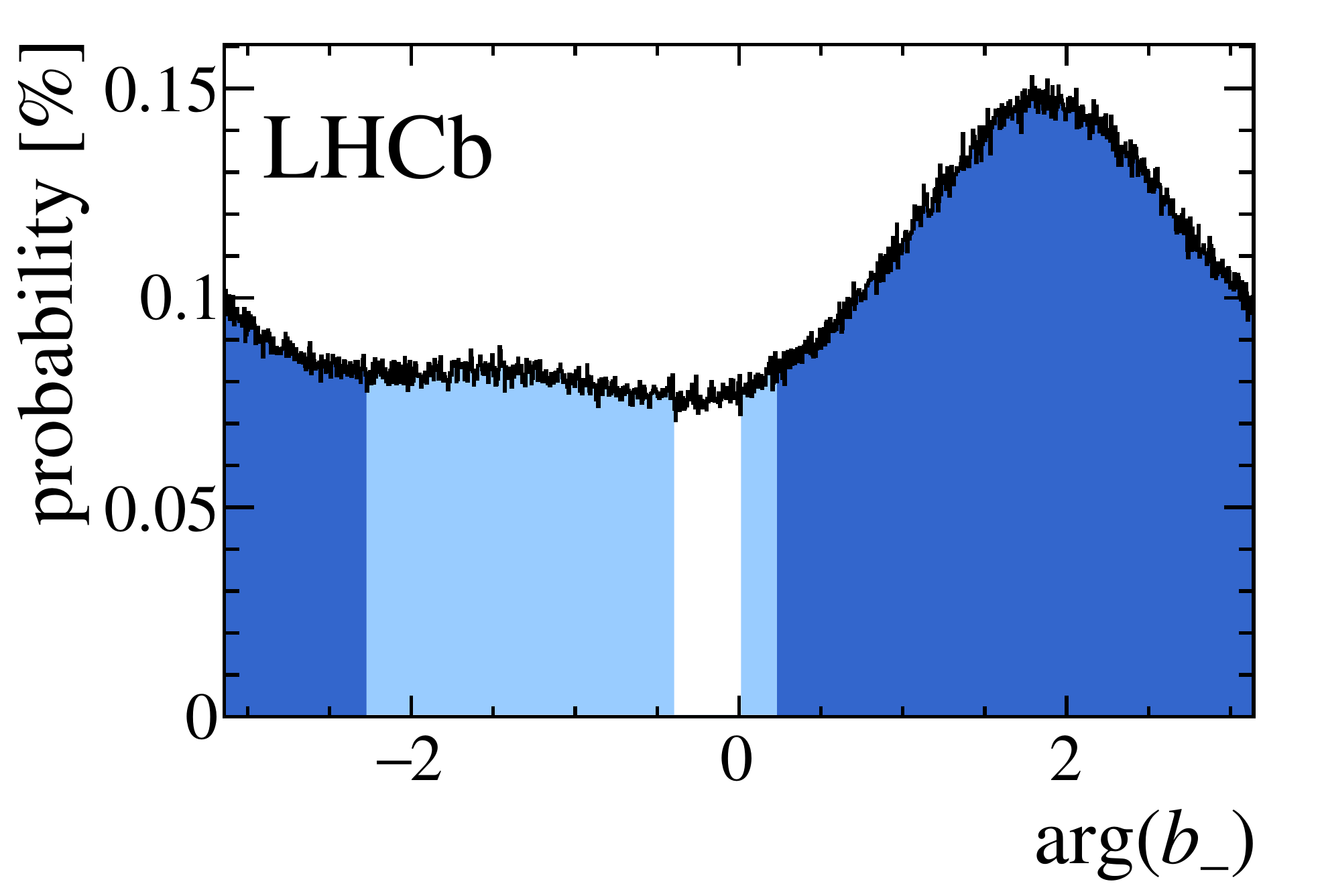} \\
    \includegraphics[width=0.32\linewidth]{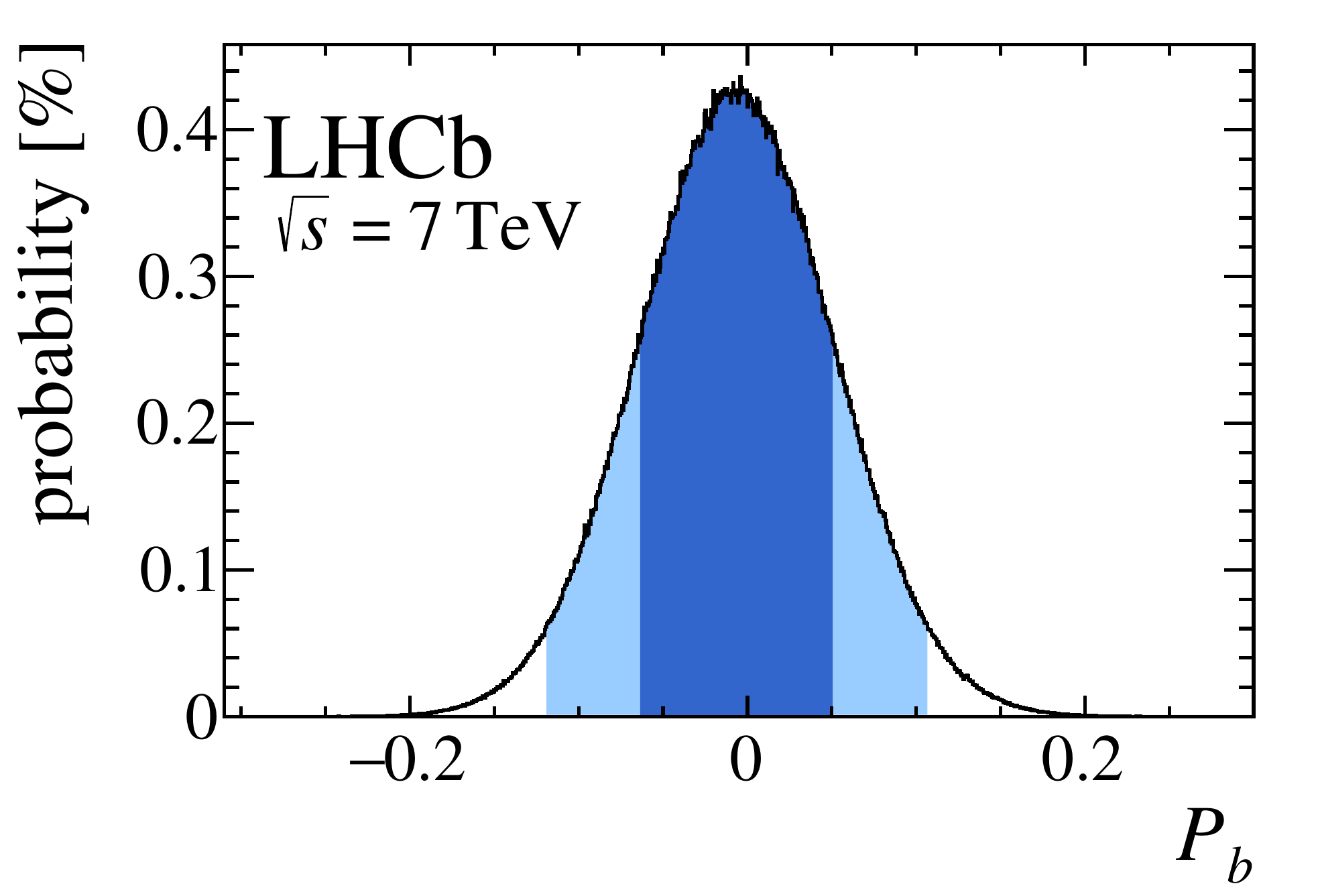} 
    \includegraphics[width=0.32\linewidth]{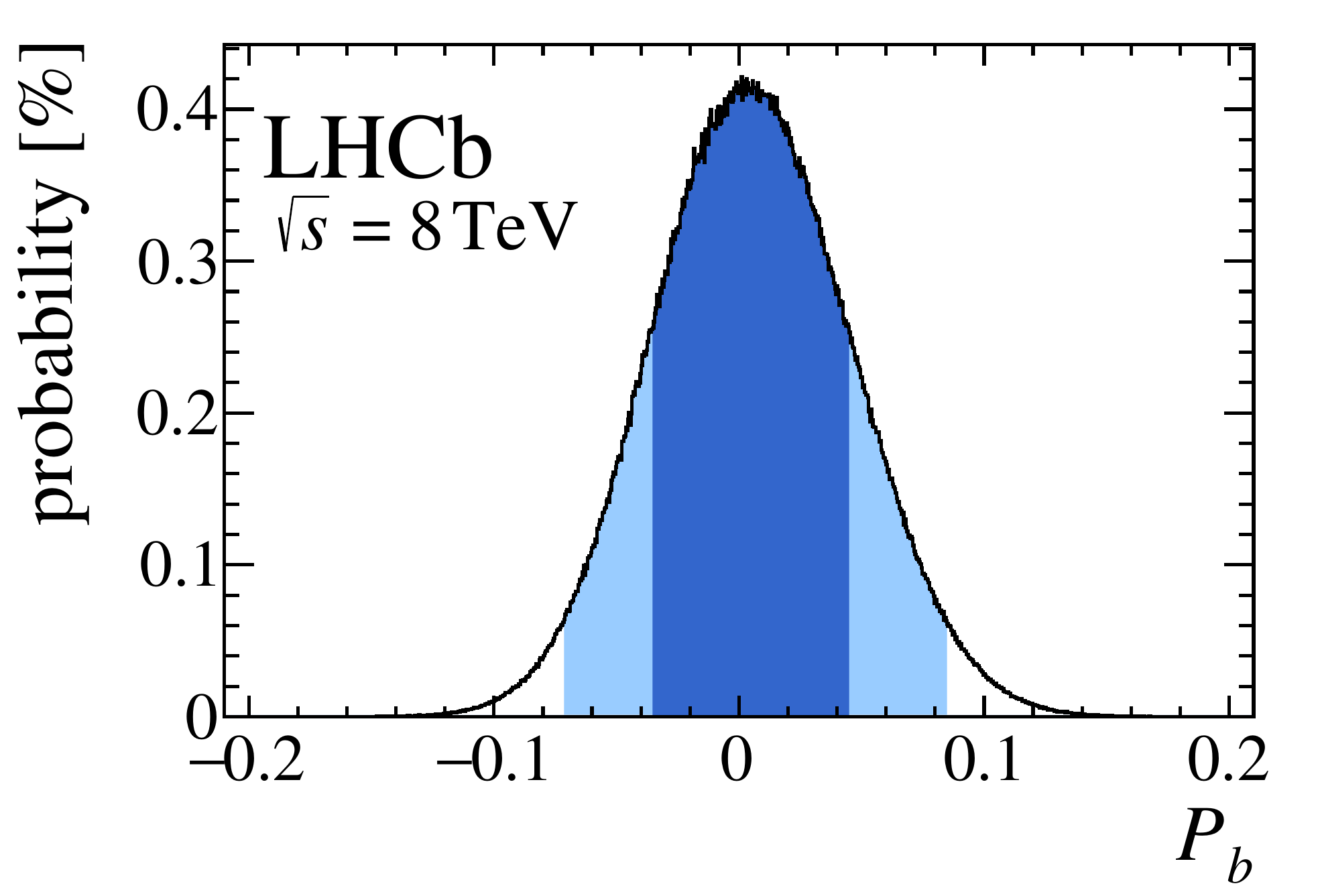} 
    \includegraphics[width=0.32\linewidth]{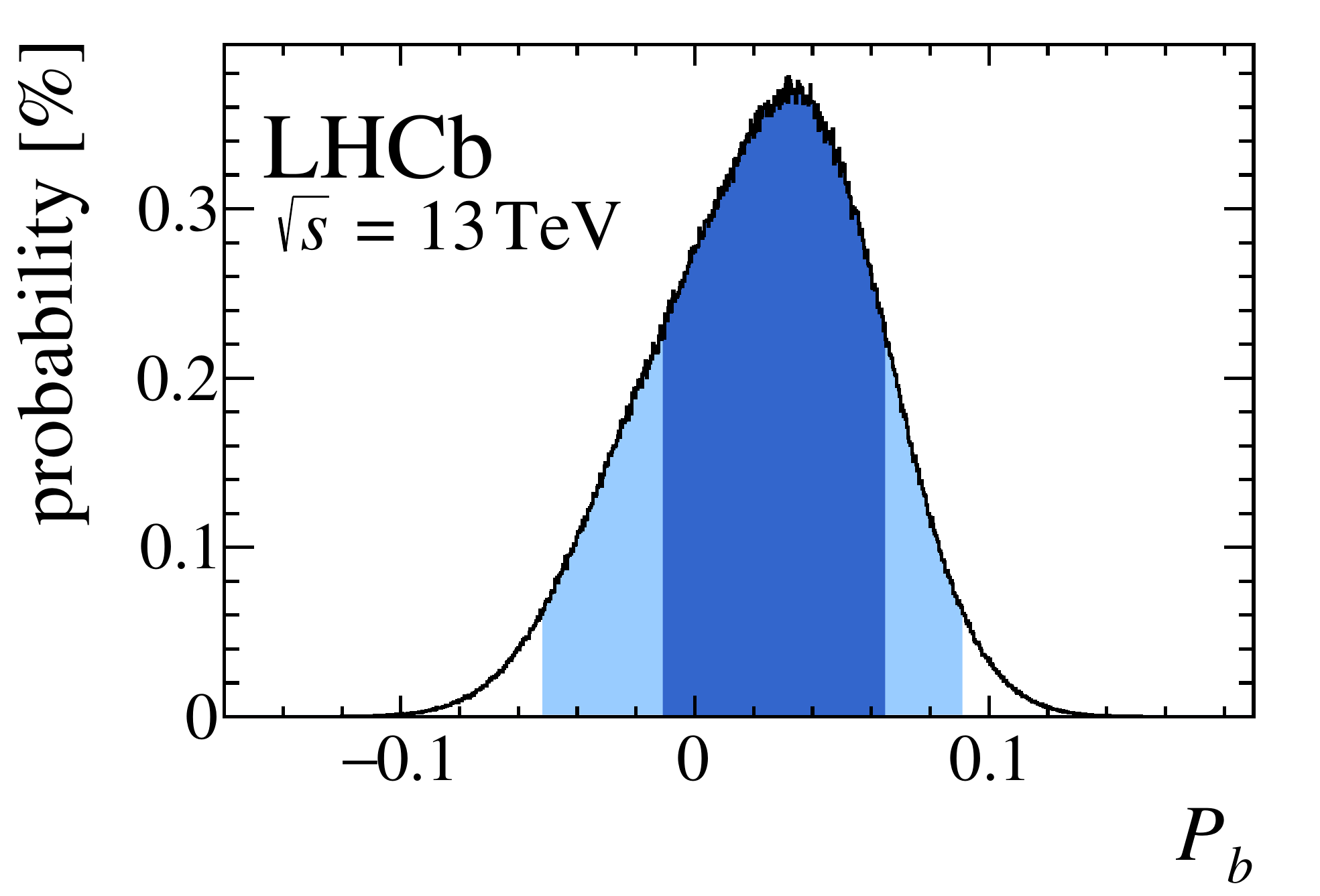} 
    \end{center}
    \caption{
    Posterior probability distributions of $|a_{\pm}|$, ${\rm arg}(a_{\pm})$, $|b_{-}|$, ${\rm arg}(b_{-})$ and the transverse production polarisation of the \Lb baryons, $P_{b}$, at centre-of-mass energies of 7, 8 and 13\tev assuming uniform priors. 
    The shaded regions indicate the 68\% and 95\% credibility intervals. 
    }
    \label{fig:results:posterior}
\end{figure}

The resulting marginal posterior distributions for the amplitudes and polarisation are shown in Figure~\ref{fig:results:posterior}. 
The magnitude and phase of $b_{+}$ are fixed to be $|b_{+}| = 1$ and ${\rm arg}(b_{+}) = 0$. 
This amplitude is one of the two amplitudes that are expected to be large. 
The remaining amplitudes are measured relative to $b_{+}$.
A uniform prior is assumed on their magnitudes and phases and on $P_b$. 
The priors use the ranges $[-1,+1]$ for $P_b$,   
$[-\pi,+\pi]$ for the phases, and the range $[0,20]$ for the magnitudes of the amplitudes. 
The values of the amplitudes and the polarisations are given in Table~\ref{tab:results:posterior}. 
The 95\% credibility intervals are provided in Table~\ref{tab:appendix:posterior} of the Appendix. 
Figure~\ref{fig:polarisation} shows $P_b$ as a function of the $\sqrt{s}$ of the data set.
The resulting \Lb polarisation at each centre-of-mass energy is found to be consistent with zero.

\begin{table}[!tb]
    \caption{
    Estimates for the magnitude and phase of the decay amplitudes and the transverse production polarisation of the \Lb baryons, extracted using the Bayesian analysis. The most probable value (MPV) and the shortest 68\% interval containing the most probable value are given.
    }
    \centering
    \begin{tabular}{crc}
    \toprule
    Observable & MPV & Interval \\
    \midrule
    $|a_{+}|$ & $\phantom{+}0.129$ & $[\phantom{+}0.033,\phantom{+}0.163]$ \\
$|a_{-}|$ & $\phantom{+}1.021$ & $[\phantom{+}0.998, \phantom{+}1.041]$ \\
$|b_{-}|$ & $\phantom{+}0.145$ & $[\phantom{+}0.060, \phantom{+}0.188]$ \\
    ${\rm arg}(a_{+})$ [rad] & $-2.523$ & $[-\pi, -1.131]~\text{or}~[2.117,\pi]$
 \\
${\rm arg}(a_{-})$ [rad] & $\phantom{+}1.122$ & $[-2.633,-1.759]~\text{or}~[0.101,2.224]$
 \\
${\rm arg}(b_{-})$ [rad] & $\phantom{+}1.788$ & $[-\pi,-2.275]~\text{or}~[0.232,\pi]$ \\
$P_{b}$ (7\tev) & $-0.004$ & $[-0.064,\phantom{+}0.051]$ \\
$P_{b}$ (8\tev) & $\phantom{+}0.001$ & $[-0.035, \phantom{+}0.045]$ \\
$P_{b}$ (13\tev) & $\phantom{+}0.032$ & $[-0.011, \phantom{+}0.065]$ \\
$\alpha_b$ & $-0.022$ & $[-0.048, \phantom{+}0.005]$ \\
    \bottomrule
    \end{tabular}
    \label{tab:results:posterior}
\end{table}

\begin{figure}[!tb]
    \centering
    \includegraphics[width=0.6\linewidth]{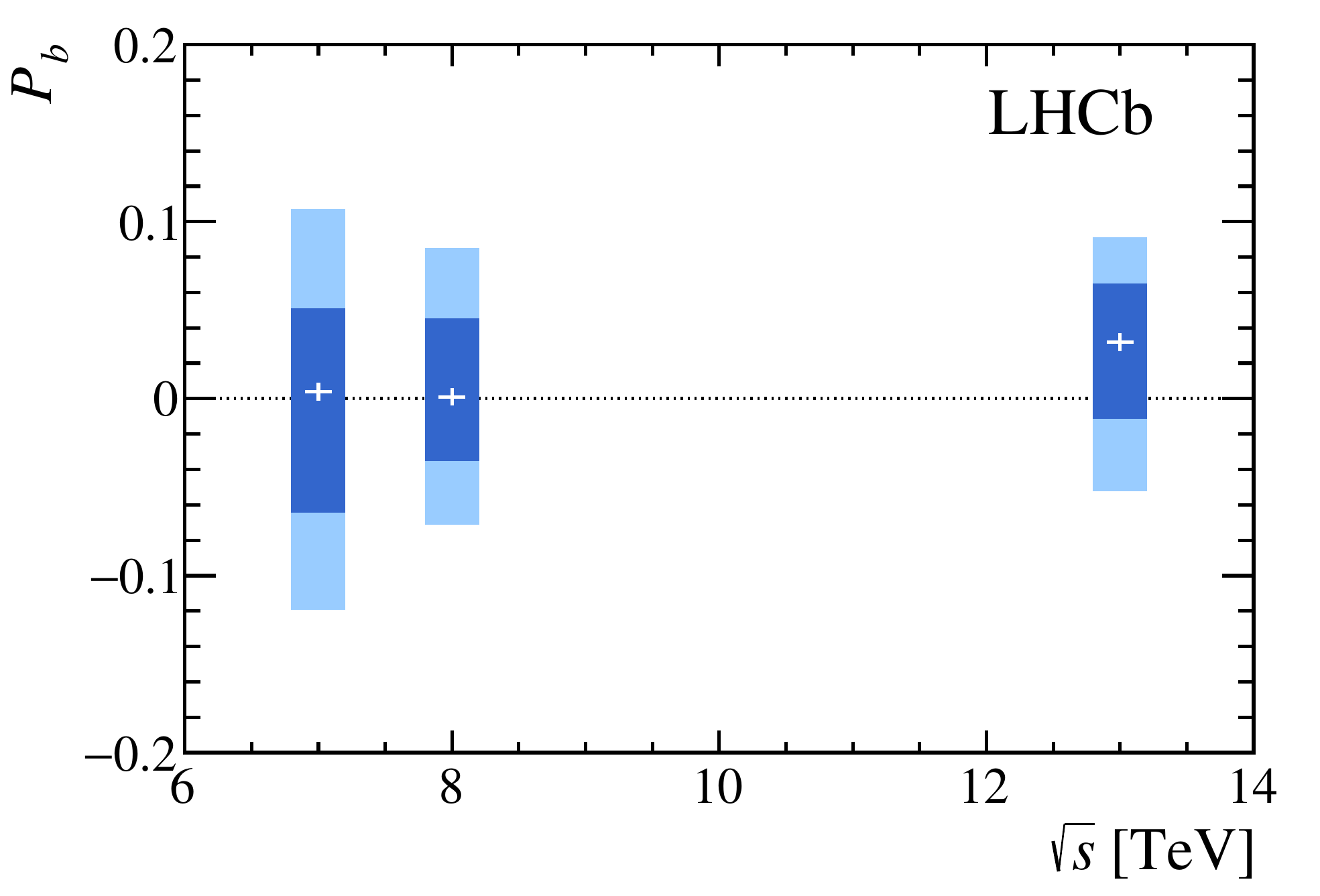}
    \caption{
    Measured transverse production polarisation of the \Lb baryons, $P_{b}$, as a function of the centre-of-mass energy, $\sqrt{s}$, of the data set.
    The points indicate the most probable value and the shaded regions the 68\% and 95\% credibility level intervals.   
    }
    \label{fig:polarisation}
\end{figure}

\begin{figure}[!htb]
    \centering
    \includegraphics[width=0.48\linewidth]{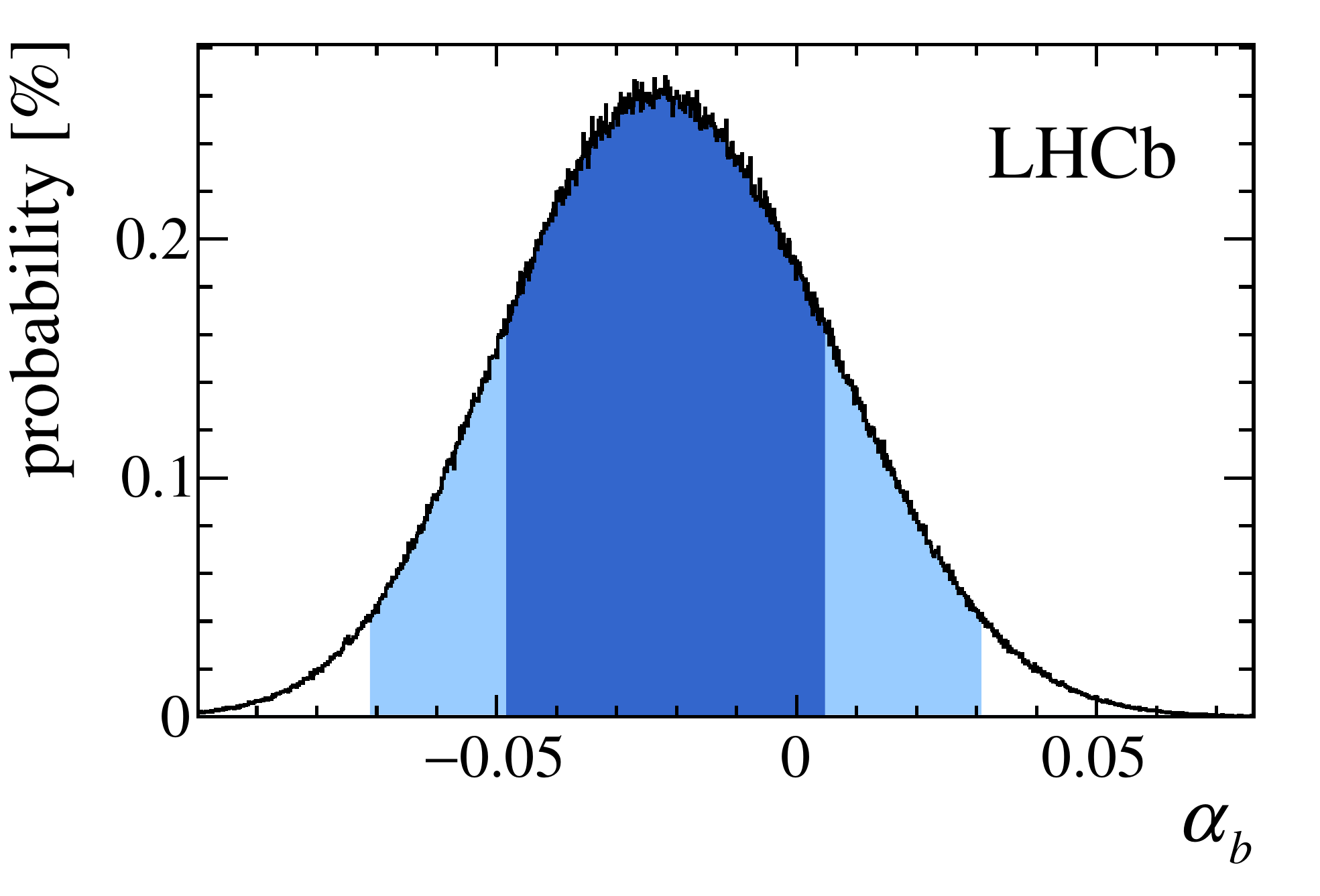}
    \caption{
    Posterior probability distribution of the parity-violating asymmetry parameter, $\alpha_b$.
    The shaded regions indicate the 68\% and 95\% credibility intervals. 
    }
    \label{fig:supp:alphab}
\end{figure}

The Markov chain finds two almost-degenerate solutions, which correspond to a change in sign of the polarisation accompanied by a change in sign of the decay amplitudes. 
This occurs due to the small size of two of the amplitudes. 
The degeneracy is most visible in the posterior distribution of $P_{b}$ determined at $\sqrt{s}$ of 13\tev, leading to an asymmetric distribution.  
Due to the small size of polarisation, there is little sensitivity to the phases of the amplitudes. 
The magnitudes of the amplitudes $a_{+}$ and $b_{-}$ are consistent with zero at the 95\% credibility level, as expected in the heavy-quark limit. 
The magnitudes of $a_{-}$ and $b_{+}$ are found to be similar in size. 
Figure~\ref{fig:supp:alphab} shows the posterior distribution of the parity-violating asymmetry parameter, $\alpha_b$, from the Bayesian analysis. 
The most probable value of $\alpha_b$ is $-0.022$. 
The 68\% credibility interval around the most probable value is $[-0.048, 0.005]$. 
This measurement is consistent with, but more precise than, previous measurements of $\alpha_b$ by the ATLAS, CMS and LHCb collaborations~\cite{Aad:2014iba,Sirunyan:2018bfd,LHCb-PAPER-2012-057}.

\begin{figure}[!tb]
    \centering
    \includegraphics[width=0.48\linewidth]{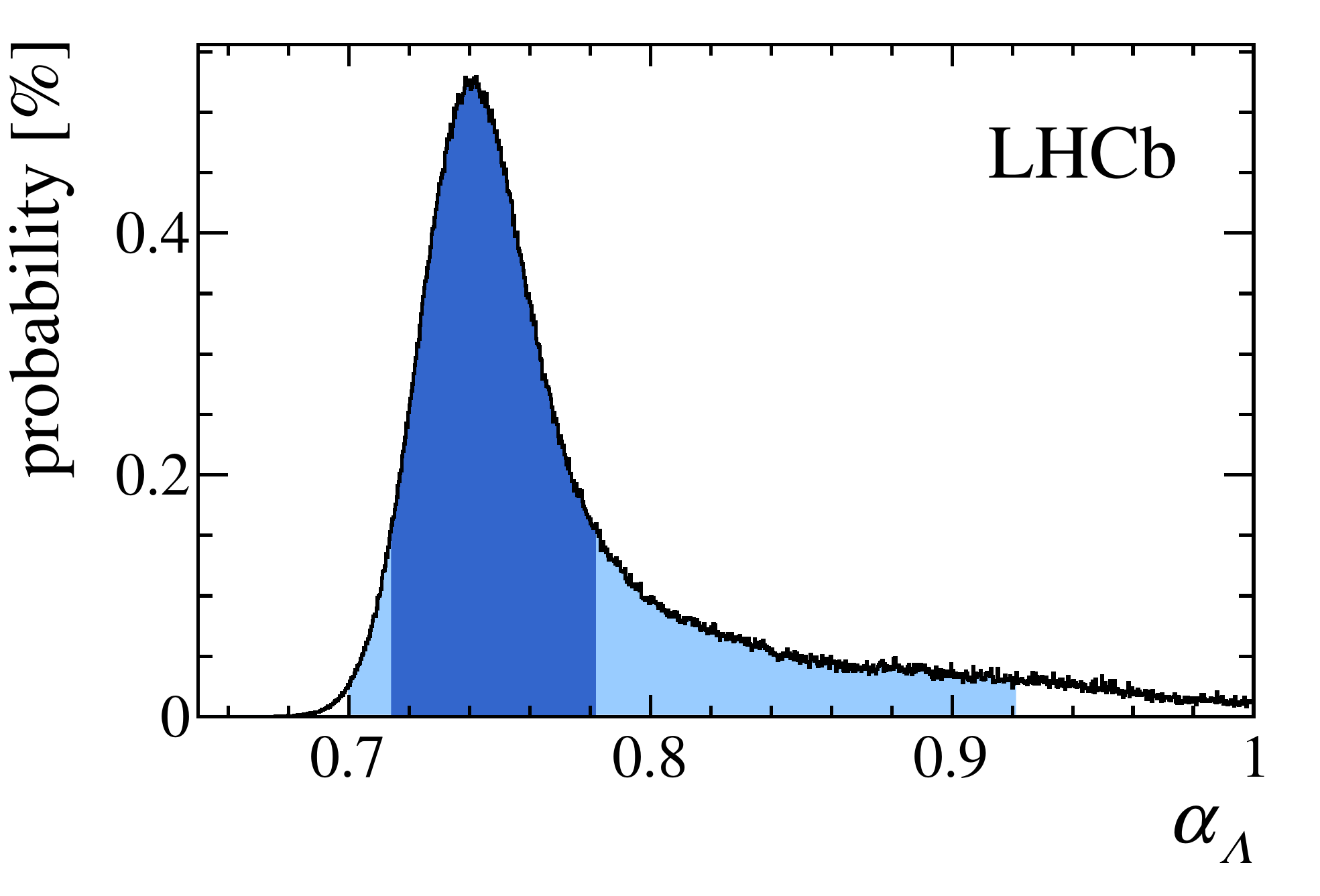}
    \caption{
    Posterior probability distribution for $\alpha_{\Lz}$, assuming a uniform prior, with all external constraints removed. 
    The shaded regions indicate the 68\% and 95\% credibility intervals. 
    }
    \label{fig:results:alpha}
\end{figure}

The posterior distribution of $\alpha_{\Lz}$ with the constraint on $\alpha_{\Lz}$ removed, assuming a uniform prior in the range $[-1,+1]$, is shown in Fig.~\ref{fig:results:alpha}. 
The most probable value of $\alpha_{\Lz}$ is 0.74. 
The 68\% credibility interval spans $[0.71, 0.78]$. 
The data strongly favour the larger $\alpha_{\Lz}$ value reported by the BES\,III collaboration~\cite{Ablikim:2018zay} over the values from older secondary scattering measurements~\cite{Astbury:1975hn,Cleland:1972fa,Dauber:1969hg,Overseth:1967zz,Cronin:1963zb}, which are excluded with high significance. 
Small values of $\alpha_{\Lz}$ are excluded by the large $\proton\pim$ forward-backward asymmetry observed in the  $\cos\theta_b$ distribution. 
Larger values of $\alpha_{\Lz}$ can be accommodated by changing the magnitudes of the decay amplitudes to reduce the asymmetry.

\section{Summary}
\label{sec:Summary}

This paper presents a measurement of the decay  amplitudes parameterising the \decay{\Lb}{\jpsi\Lz} angular distribution, and a measurement of the transverse production polarisation of the \Lb baryons at $\sqrt{s}$ of 7, 8 and 13\tev, using data collected with the LHCb experiment. 
The measurements are performed in a fiducial region of \Lb transverse momentum and pseudorapidity of $1 < \pt < 20\gevc$ and $2 < \eta < 5$, respectively.
The magnitudes of two of the four decay amplitudes are found to be small. One of these amplitudes corresponds to \Lz helicity of $+\tfrac{1}{2}$ and \jpsi helicity of $0$ and the other to \Lz helicity of $-\tfrac{1}{2}$ and \jpsi helicity of $-1$. 
This is consistent with the expectation from the heavy-quark limit and the left-handed nature of the weak interaction.
The parity-violating parameter $\alpha_b$ is found to be consistent with zero, with a 68\% credibility interval from $-0.048$ to $0.005$.
The small negative value of $\alpha_b$ favoured by the data is consistent with most theoretical predictions but is inconsistent with the prediction based on HQET in Ref.~\cite{Ajaltouni:2004zu}.
The \Lb production polarisation is found to be consistent with zero, with 68\% credibility level intervals of $[-0.06,0.05]$, $[-0.04,0.05]$ and $[-0.01,0.07]$ at $\sqrt{s}$ of 7, 8 and 13\tev, respectively. 
The results in this paper supersede those of  Ref.~\cite{LHCb-PAPER-2012-057} and are largely consistent with the previous measurements~\cite{LHCb-PAPER-2012-057,Sirunyan:2018bfd,Aad:2014iba}.
Differences between the results presented in this paper and the previous measurements can be attributed to the value of $\alpha_{\Lz}$ used in those measurements. 
The data strongly support the recent BES\,III measurement of $\alpha_{\Lz}$ over the previous value from secondary scattering data. 
With the old value of $\alpha_{\Lz}$, it is not possible to describe the data with a physical set of amplitudes.

\section*{Acknowledgements}
%
%
\noindent We express our gratitude to our colleagues in the CERN
accelerator departments for the excellent performance of the LHC. We
thank the technical and administrative staff at the LHCb
institutes.
We acknowledge support from CERN and from the national agencies:
CAPES, CNPq, FAPERJ and FINEP (Brazil); 
MOST and NSFC (China); 
CNRS/IN2P3 (France); 
BMBF, DFG and MPG (Germany); 
INFN (Italy); 
NWO (Netherlands); 
MNiSW and NCN (Poland); 
MEN/IFA (Romania); 
MSHE (Russia); 
MinECo (Spain); 
SNSF and SER (Switzerland); 
NASU (Ukraine); 
STFC (United Kingdom); 
DOE NP and NSF (USA).
We acknowledge the computing resources that are provided by CERN, IN2P3
(France), KIT and DESY (Germany), INFN (Italy), SURF (Netherlands),
PIC (Spain), GridPP (United Kingdom), RRCKI and Yandex
LLC (Russia), CSCS (Switzerland), IFIN-HH (Romania), CBPF (Brazil),
PL-GRID (Poland) and OSC (USA).
We are indebted to the communities behind the multiple open-source
software packages on which we depend.
Individual groups or members have received support from
AvH Foundation (Germany);
EPLANET, Marie Sk\l{}odowska-Curie Actions and ERC (European Union);
ANR, Labex P2IO and OCEVU, and R\'{e}gion Auvergne-Rh\^{o}ne-Alpes (France);
Key Research Program of Frontier Sciences of CAS, CAS PIFI, and the Thousand Talents Program (China);
RFBR, RSF and Yandex LLC (Russia);
GVA, XuntaGal and GENCAT (Spain);
the Royal Society
and the Leverhulme Trust (United Kingdom).

\clearpage

\section*{Appendices}

\appendix

\section{Correlation matrices}
\label{sec:appendix:correlation}

\begin{figure}[!htb]
    \centering
    \includegraphics[width=0.55\linewidth]{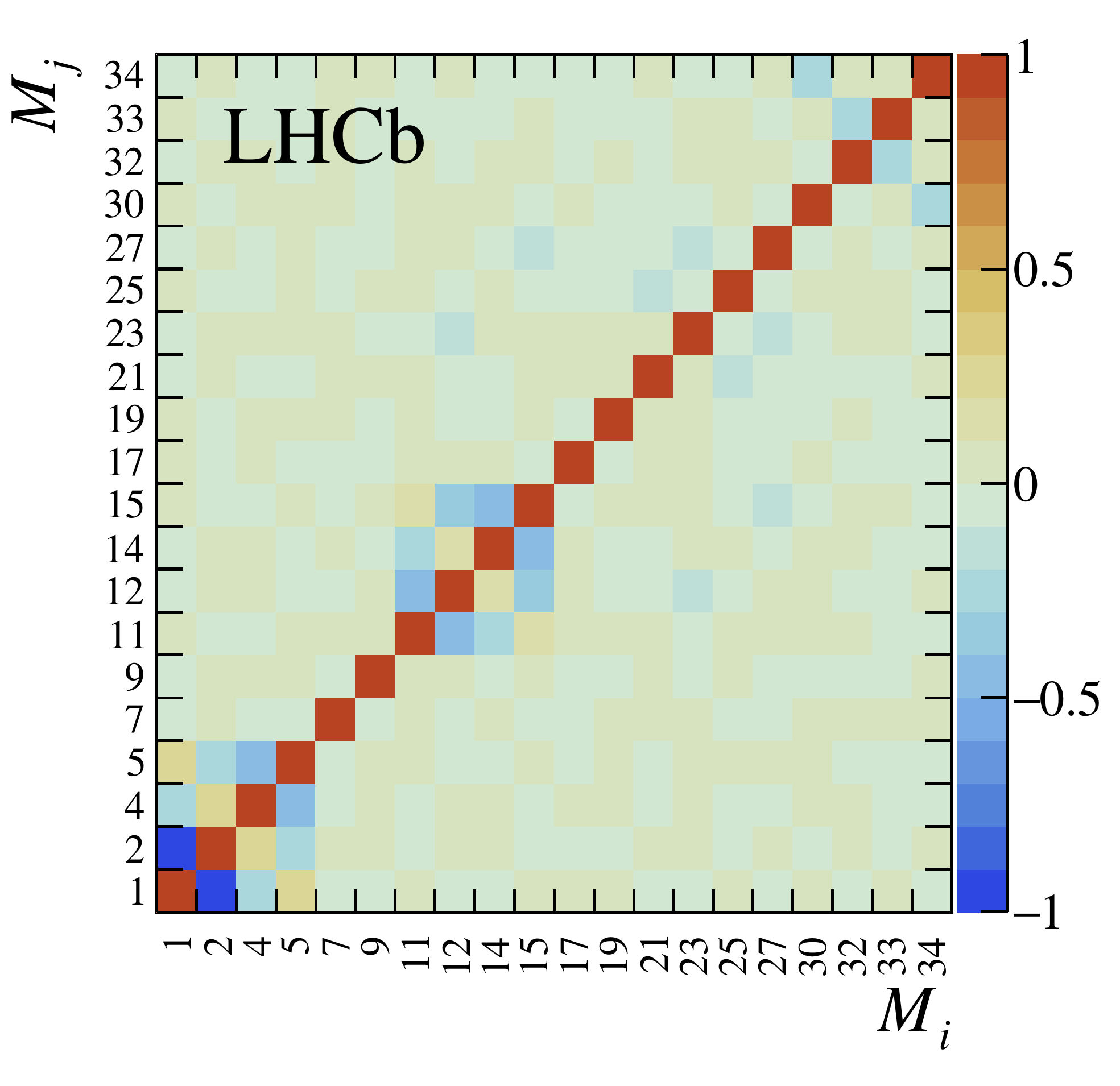}
    \caption{Statistical correlation between the moments determined at $\sqrt{s}$ of 7\tev.}
    \label{fig:correlation:7TeV}
\end{figure}

\begin{figure}[!htb]
    \centering
    \includegraphics[width=0.55\linewidth]{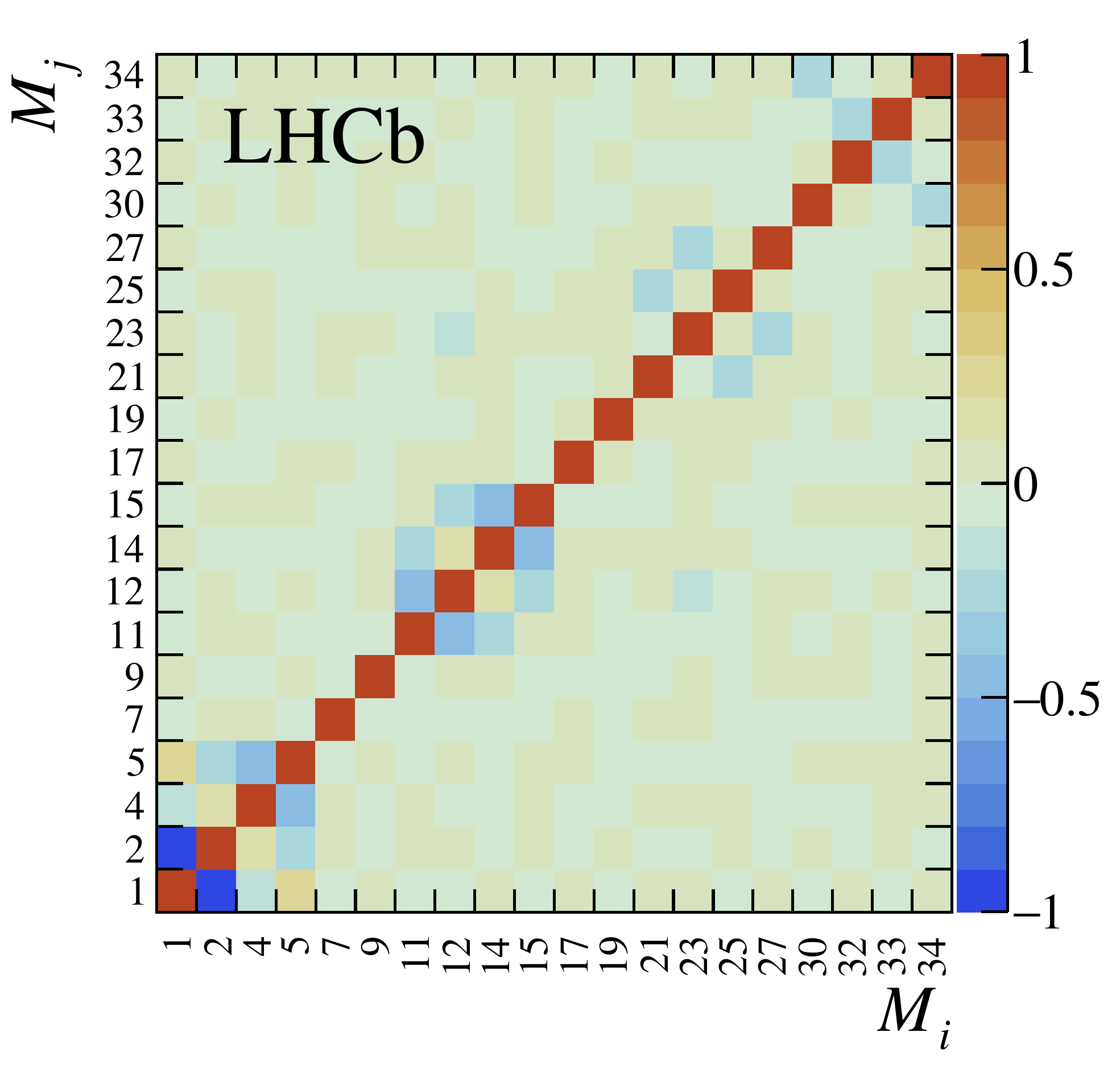}
    \caption{Statistical correlation between the moments determined at $\sqrt{s}$ of 8\tev.}
    \label{fig:correlation:8TeV}
\end{figure}

\begin{figure}[!htb]
    \centering
    \includegraphics[width=0.55\linewidth]{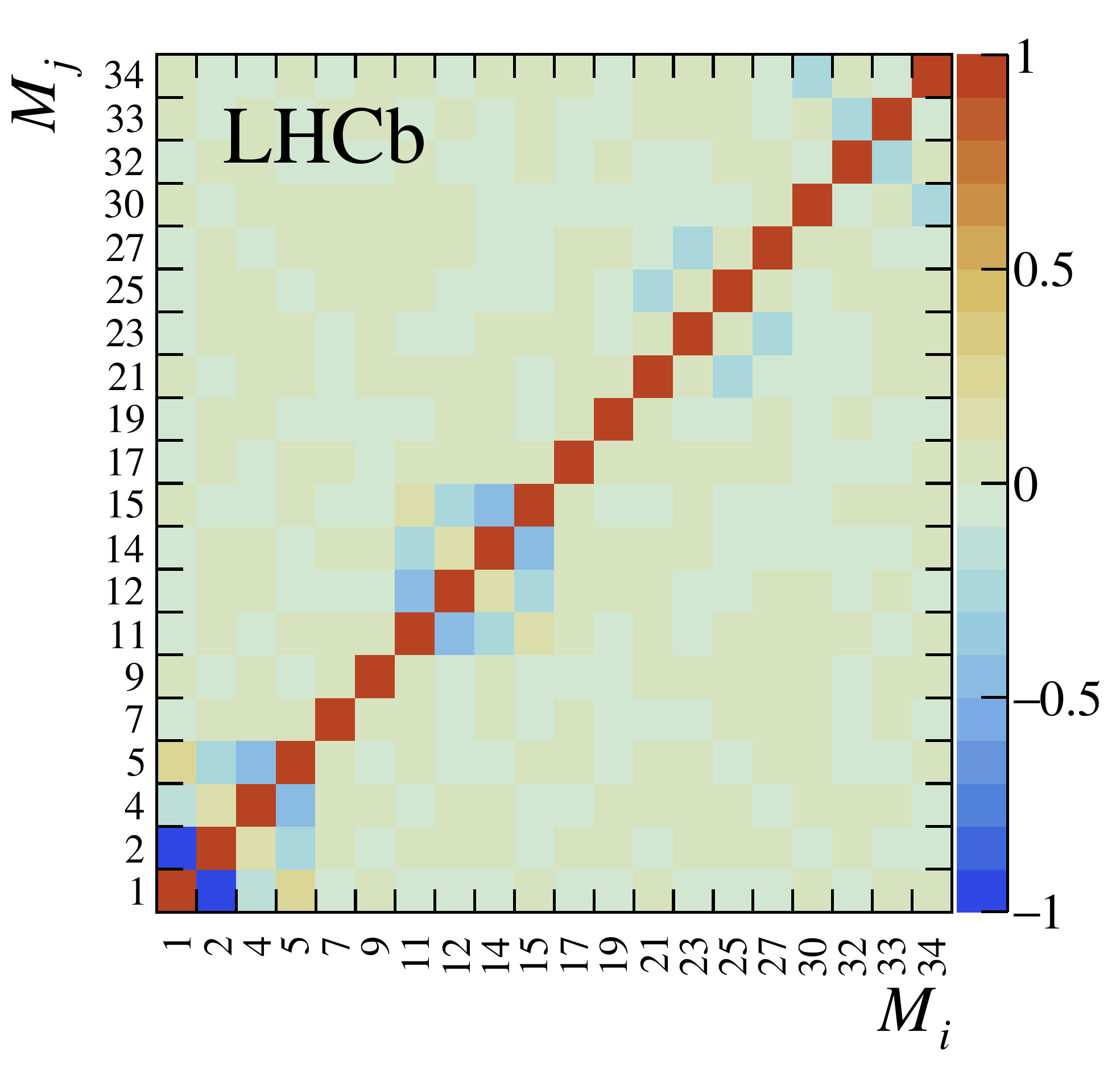}
    \caption{Statistical correlation between the moments determined at $\sqrt{s}$ of 13\tev.}
    \label{fig:correlation:13TeV}
\end{figure}

The statistical correlations between the different moments determined at the three different centre-of-mass energies are shown in Figs.~\ref{fig:correlation:7TeV}, \ref{fig:correlation:8TeV} and \ref{fig:correlation:13TeV}. 
The correlation coefficients are determined by bootstrapping the data set. 
The covariance matrices are available as supplementary material to this article.

\clearpage

\section{Intervals at 95\% credibility level}
\label{sec:appendix:intervals}

The 95\% credibility level intervals on the decay amplitudes and production polarisation from the Bayesian analysis of the moments are given in Table~\ref{tab:appendix:posterior}. The 95\% intervals on $\alpha_b$ and on $\alpha_\Lz$ are also provided. The interval on $\alpha_\Lz$ is evaluated after removing the external constraint on that parameter. 

\begin{table}[!htb]
    \caption{
    Intervals at 95\% credibility level on the amplitudes, the polarisation and $\alpha_{b}$ from the Bayesian analysis. 
    The interval on $\alpha_{\Lz}$, with the external constraint removed, is also provided. 
    }
    \centering
    \begin{tabular}{cc}
        \toprule
        Observable & Interval \\
        \midrule
    $|a_{+}|$ & $[\phantom{+}0.000,\phantom{+}0.200]$ \\
    $|a_{-}|$ & $[\phantom{+}0.978,\phantom{+}1.063]$ \\
    $|b_{-}|$ & $[\phantom{+}0.000,\phantom{+}0.208]$ \\
    ${\rm arg}(a_{+})$ [rad] &  $[-\pi, \phantom{+}0.251]~\text{or}~[\phantom{+}0.848,\pi]$ \\
${\rm arg}(a_{-})$ [rad] &  $[-\pi,-1.137]~\text{or}~[-0.459,\pi]$
 \\
${\rm arg}(b_{-})$ [rad] &   $[-\pi,-0.396]~\text{or}~[\phantom{+}0.013,\pi]$
\\
$P_{b}$ (7\tev) & $[-0.119,\phantom{+}0.107]$ \\
$P_{b}$ (8\tev) & $[-0.071,\phantom{+}0.085]$ \\
$P_{b}$ (13\tev) & $[-0.052,\phantom{+}0.091]$ \\
$\alpha_b$ & $[-0.071,\phantom{+}0.031]$ \\
$\alpha_\Lz$ & $[\phantom{+}0.700,\phantom{+}0.921]$ \\
\bottomrule
    \end{tabular}
    \label{tab:appendix:posterior}
\end{table}

\addcontentsline{toc}{section}{References}
\bibliographystyle{LHCb}
\bibliography{main,standard,LHCb-PAPER,LHCb-CONF,LHCb-DP,LHCb-TDR}

\newpage
\centerline
{\large\bf LHCb collaboration}
\begin
{flushleft}
\small
R.~Aaij$^{31}$,
C.~Abell{\'a}n~Beteta$^{49}$,
T.~Ackernley$^{59}$,
B.~Adeva$^{45}$,
M.~Adinolfi$^{53}$,
H.~Afsharnia$^{9}$,
C.A.~Aidala$^{81}$,
S.~Aiola$^{25}$,
Z.~Ajaltouni$^{9}$,
S.~Akar$^{66}$
% ...

{\footnotesize \it

% ...
$ ^{9}$Universit{\'e} Clermont Auvergne, CNRS/IN2P3, LPC, Clermont-Ferrand, France\\
% ...
$ ^{25}$INFN Sezione di Milano, Milano, Italy\\
% ...
$ ^{31}$Nikhef National Institute for Subatomic Physics, Amsterdam, Netherlands\\
% ...
$ ^{45}$Instituto Galego de F{\'\i}sica de Altas Enerx{\'\i}as (IGFAE), Universidade de Santiago de Compostela, Santiago de Compostela, Spain\\
% ...
$ ^{49}$Physik-Institut, Universit{\"a}t Z{\"u}rich, Z{\"u}rich, Switzerland\\
% ...
$ ^{53}$H.H. Wills Physics Laboratory, University of Bristol, Bristol, United Kingdom\\
% ...
$ ^{59}$Oliver Lodge Laboratory, University of Liverpool, Liverpool, United Kingdom\\
% ...
$ ^{66}$Los Alamos National Laboratory (LANL), Los Alamos, United States\\
$ ^{67}$Syracuse University, Syracuse, NY, United States\\
% ...
$ ^{81}$University of Michigan, Ann Arbor, United States, associated to $^{67}$\\
\bigskip
% ...
\medskip
}
\end{flushleft}

\end{document}



\pagestyle{plain} 
\setcounter{page}{1}
\pagenumbering{arabic}

\clearpage

\section*{LHCb-PAPER-2020-005 supplementary material}
\label{sec:Supplementary-App}

The angular efficiency of the 2016 data set is shown in Fig.~\ref{fig:supp:eff}. 
Similar distributions are seen for the other data sets used in this analysis. 
The long and downstream efficiencies differ due to differences in the detector acceptance between the two classes of candidates. 
The invariant mass and angular distributions of selected \decay{\Bz}{\jpsi\KS} decays are shown in Fig.~\ref{fig:supp:control}. 
The mass fit is used to background subtract the data. 
The background-subtracted angular distribution is well described by the product of 
\begin{align}
    \frac{\deriv^{5}\Gamma[\decay{\Bz}{\jpsi\KS}]}{\deriv \vec{\Omega}} \propto \sin^2\theta_{l}
    \label{eq:control}
\end{align}
and the detector efficiency derived by simulation. 
Equation~\ref{eq:control} is equivalent to setting $M_{1} = \tfrac{1}{2}$ and $M_{2}$--$M_{34}$ equal to zero in the angular distribution defined by Equation~$2$ in the paper.
Figure~\ref{fig:supp:masses} shows the $\proton\pim$ and the \mumu mass of selected \decay{\Lb}{\jpsi\Lz} candidates.

\begin{figure}[!htb]
    \centering
    \includegraphics[width=0.33\linewidth]{figs/Fig1a-supp.pdf} 
    \includegraphics[width=0.33\linewidth]{figs/Fig1b-supp.pdf} \\
    \includegraphics[width=0.33\linewidth]{figs/Fig1c-supp.pdf} 
    \includegraphics[width=0.33\linewidth]{figs/Fig1d-supp.pdf} \\
    \includegraphics[width=0.33\linewidth]{figs/Fig1e-supp.pdf} 
    \includegraphics[width=0.33\linewidth]{figs/Fig1f-supp.pdf} \\
    \includegraphics[width=0.33\linewidth]{figs/Fig1g-supp.pdf} 
    \includegraphics[width=0.33\linewidth]{figs/Fig1h-supp.pdf} \\
    \includegraphics[width=0.33\linewidth]{figs/Fig1i-supp.pdf} 
    \includegraphics[width=0.33\linewidth]{figs/Fig1j-supp.pdf} 
    \caption{
    Angular projections of simulated phase-space decays, compared to the result of the efficiency model for (left) long and (right) downstream candidates in the 2016 data set. 
    The data points represent the simulated candidates and the line represents the efficiency model. 
    }
    \label{fig:supp:eff}
\end{figure}

\begin{figure}[!htb]
    \begin{center}
    \includegraphics[width=0.48\linewidth]{figs/Fig2a-supp.pdf}
    \includegraphics[width=0.48\linewidth]{figs/Fig2b-supp.pdf} \\
    \includegraphics[width=0.48\linewidth]{figs/Fig2c-supp.pdf}
    \includegraphics[width=0.48\linewidth]{figs/Fig2d-supp.pdf} \\
    \includegraphics[width=0.48\linewidth]{figs/Fig2e-supp.pdf}
    \includegraphics[width=0.48\linewidth]{figs/Fig2f-supp.pdf}
    \end{center}
    \caption{
    Invariant mass and angular distributions of selected \decay{\Bz}{\jpsi\KS} decays. 
    The long and downstream categories and the different data taking years have been combined. 
    The angular distribution has been background-subtracted and is compared to the  result of the moment analysis, folded with the angular efficiency.
    }
    \label{fig:supp:control}
\end{figure}

\begin{figure}[!htb]
    \centering
    \includegraphics[width=0.48\linewidth]{figs/Fig3a-supp.pdf}
    \includegraphics[width=0.48\linewidth]{figs/Fig3b-supp.pdf}
    \caption{
    The (left) $\proton\pim$ and (right) \mumu mass of selected \decay{\Lb}{\jpsi\Lz} candidates before the application of the multivariate selection. 
    Candidates in the long and downstream categories have been combined. 
    }
    \label{fig:supp:masses}
\end{figure}

\clearpage